\documentclass[12pt]{iopart}
\usepackage{epsfig}
\usepackage{graphicx} 
\usepackage{wrapfig} 
\usepackage[dvips]{color}

\begin{document}

\newcommand{\lesssim}   {\mathrel{\mathop{\kern 0pt \rlap
  {\raise.2ex\hbox{$<$}}}
  \lower.9ex\hbox{\kern-.190em $\sim$}}}
\newcommand{\gsim}   {\mathrel{\mathop{\kern 0pt \rlap
  {\raise.2ex\hbox{$>$}}}
  \lower.9ex\hbox{\kern-.190em $\sim$}}}

\newcommand{\be}{\begin{equation}}
\newcommand{\ee}{\end{equation}}
\newcommand{\ba}{\begin{eqnarray}}
\newcommand{\ea}{\end{eqnarray}}
\def\bone{$B^{(1)}$}
\def\bone{B^{(1)}}
\def\etal{{\it et al.~}}
\def\eg{{\it e.g.~}}
\def\ie{{\it i.e.~}}
\def\DM{dark matter~}
\def\DE{dark energy~} 
\def\GC{Galactic center~} 
\def\susy{SUSY~}

{\hfill CERN-PH-TH/2007-161}

\title{Composition of UHECR and the Pierre Auger Observatory Spectrum}

\author{Katsushi Arisaka$^a$, Graciela B. Gelmini$^{a,b}$, Matthew Healy$^a$, Oleg Kalashev$^{c}$ and Joong Lee$^a$}

\address{$^a$ Department of Physics and Astronomy, UCLA, Los Angeles,
CA 90095-1547, USA}
\address{$^b$ CERN, PH-TH, CH-1211 Gen\`eve 23, Switzerland}
\address{$^c$INR RAS, 60th October 
Anniversary pr. 7a, 117312 Moscow, Russia}

\begin{abstract}
 We fit the recently published Pierre Auger ultra-high energy cosmic ray spectrum assuming that either nucleons or nuclei are emitted at the sources. We consider the simplified cases of pure proton, or pure oxygen, or pure iron injection.  We perform an exhaustive scan in the source evolution factor, the spectral index, the maximum energy of the source spectrum $Z\times E_{\rm max}$, and the minimum distance to the sources. We show that the Pierre Auger spectrum agrees with any of the source compositions we assumed. For iron, in particular, there are two distinct solutions  with high and low $E_{\rm max}$ 
(e.g.  $6.4 \times 10^{20}$~eV and $2 \times 10^{19}$~eV) respectively which could be distinguished by either a large fraction or the near absence of proton primaries at the highest energies.
  We raise the possibility that an iron dominated injected flux  may be in line with the latest composition measurement from the Pierre Auger Observatory where a hint of  heavy element dominance is seen.
\end{abstract}

\maketitle

\section{Introduction}

The Greisen-Zatsepin-Kuzmin (GZK) cutoff~\cite{gzk} at $4 \times 10^{19}$~eV seems not to be present in the data of the AGASA ground array~\cite{agasa} but it appears in the data of the HiRes 
air fluorescence detector~\cite{hires, hires_mono_spec}.  This controversy can be addressed by the Pierre Auger Observatory~\cite{Auger}, a hybrid combination of charged particle detectors
and fluorescence telescopes, as it continues to accumulate data. We study here the most recent spectrum published  by the Pierre Auger Observatory~\cite{ICRC2007-auger-energy-spectrum}.

Using the surface array, the Pierre Auger Collaboration 
presented~\cite{ICRC2007-auger-energy-spectrum} an
update to their previous result~\cite{ICRC2005-auger-energy-spectrum}
that includes two additional years of data and an integrated
aperture (5165~km$^2$~sr~yr) nearly equivalent to that of the HiRes
experiment.  The updated spectrum begins at an energy of
$2.5 \times 10^{18}$~eV, the energy at which the surface array becomes
fully efficient within the zenith angle range
0-60$^\circ$~\cite{ICRC2005-auger-acceptance}, and ends with a
highest observed energy of $\sim 1.8\times 10^{20}$~eV.  Energies are
determined in a simulation independent way assuming constant intensity
and calibrating the ground observable S(1000) against the fluorescence
detector energy for the subset of showers (known as golden hybrid
showers) that contain reconstructions from both detectors.  The method
leads to a statistical error of 8\% and a systematic error of
22\%~\cite{ICRC2007-auger-energy-spectrum, ICRC2007-auger-energy-determination}
on the energy.

The origin of cosmic rays with energies beyond the GZK
cutoff remain an outstanding open question
 in astroparticle physics and  cosmology~\cite{agasa, hires, hires_mono_spec, agasa_spec}. Nucleons
cannot be significantly deflected by the magnetic fields
  of our  galaxy for energies above the ``ankle", i.e.  above
10$^{18.5}$~eV. This and the absence of a correlation of arrival directions
with the galactic plane indicate that, if nucleons are the primary particles
of the ultra high energy cosmic rays (UHECR), these nucleons should be of
extragalactic origin. Moreover, nucleons  as well as photons with energies above 
$5 \times 10^{19}$~eV could not reach Earth from a distance beyond 50 to 100
Mpc~\cite{50Mpc, 40Mpc} thus  sources should be found within this distance.
 Nucleons scatter off the cosmic microwave background
(CMB) photons with a resonant photoproduction of pions $p\gamma \rightarrow
\Delta^* \rightarrow N\pi$, where the pion carries away $\sim 20\%$ of the
original nucleon energy.  Photons with comparable energy pair-produce electrons and positrons
on the radio background. 

Intervening sheets of large scale intense extra galactic magnetic fields
(EGMF), with intensities $B \sim 0.1 -1\times 10^{-6}$~G, could provide
sufficient angular deflection for protons to explain the lack of observed
sources in the directions of arrival of UHECR. However, recent realistic
simulations of the expected large scale EGMF show that strong deflections
could only occur when particles cross galaxy clusters. Except in the regions
close to the Virgo, Perseus and Coma clusters the magnetic fields are
not larger than $3 \times 10^{-11}$~G~\cite{dolag2004} and the deflections
expected are not important (however  see Ref.~\cite{Sigl:2004yk}).

Heavy nuclei are an interesting possibility for UHECR primaries, since they
could be produced at the sources with larger maximum energies 
and would more easily be deflected by intervening magnetic
fields. Both AGASA and HiRes data favor 
a dominance of light hadrons, consistent with
being all protons, in the composition of UHECR above 10$^{19}$~eV~\cite{ICRC2007-hires-elongation}.  These data are consistent with models in
which  all UHECR above $10^{18}$ eV are due to 
extragalactic protons~\cite{berezinsky2002}.  The Pierre Auger
Observatory has presented an elongation rate that is better
represented by a fit containing a break point in the slope at
$2 \times 10^{18}$~eV. 
Below the break point the spectrum is consistent with a progressively
lighter composition but above the break the composition is consistent
with a constant and mixed composition up to the highest
energies~\cite{ICRC2007-auger-composition}. This raises the possibility of a
significant fraction of heavier elements in the range of the GZK cutoff.

Whether particles can be emitted with the necessary energies by astrophysical
accelerators, such as active galactic nuclei, jets or extended lobes of radio galaxies, 
or even extended objects such as colliding galaxies and
clusters of galaxies, is still an open question. The size and possible
magnetic and electric fields of these astrophysical sites make it plausible
for them to accelerate protons  and nuclei  to a maximum energy of
$Z \times10^{21}$~eV, where $Z$ is the number of protons in each
nucleus.  Larger emission
energies would require a reconsideration of possible acceleration models or
sites.

A galactic component of the UHECR flux, which could be important 
up to energies  10$^{19}$~eV, should consist of heavy nuclei, given the
lack of correlation with the galactic plane of events at this energy (outside the
galactic plane galactic protons would be deflected by a maximum of 15-20$^o$
at this energy~\cite{galactic_magn_field}).  

 In this paper we fit the Pierre Auger UHECR spectrum  above the energy $E_{\rm cut}= 1 \times 10^{19}$~eV (and for comparison we also use two other values of 
 $E_{\rm cut}$,  $2.5 \times 10^{18}$~eV and $4 \times 10^{19}$~eV)
assuming that either protons or nuclei are emitted at the sources.
The UHECR spectrum predicted depends on the slope and maximum energy of
the  nucleon or nucleus spectrum emitted at the source, the distribution of sources, and the
 intervening backgrounds.
We take a phenomenological approach in choosing the range of the several relevant
parameters which determine the cosmic ray flux, namely we take for each 
of them a range of values mentioned in the literature, 
without attempting to assign them to particular sources or acceleration mechanisms.
We consider the simplified case in which either only protons, or only  oxygen nuclei,
or  only iron nuclei would be  emitted by the sources.  Although these are not
realistic models for the injected composition, we expect to gain
some understanding of how well a heavy or  intermediate or  light elements dominated composition 
in the injected spectrum   can account for the observed spectrum.  
     
The ankle in the UHECR spectrum at energies $10^{18}$eV - $10^{19}$ eV 
can be explained either by $e^\pm$ pair  production by  extragalactic protons interacting with the CMB~\cite{berezinsky2002} or by a
change from one component of the UHECR spectrum to another. We take into account the
first possibility by fitting the Pierre Auger spectrum above  $2.5\times 10^{18}$~eV  with
a flux of protons emitted at the sources. This  possibility can  still  be consistent
 with the proton-dominated composition observed by HiRes. 

The second explanation of the ankle, in which the extragalactic component dominates at energies above the ankle, assumes the existence of a low energy component (LEC) 
when necessary to fit the UHECR spectrum at energies lower than  1 to
$4 \times 10^{19}$~eV.   This LEC 
can be dominated by galactic Fe or by a different population of lower
energy extragalactic nucleons. Here we do not address
the issue of what the  LEC is.  We only assume that, if it exists,  it becomes negligible
at energies above  the energy $E_{\rm cut}$ at which we start our fit, i.e. either
$1\times 10^{19}$~eV  or $4\times 10^{19}$~eV. In this case we study
both the case of protons as well as that of nuclei (Fe or O) emitted at the
sources.
  
Our calculations do not take into account deflections. Since we assume  typical extragalactic magnetic fields  not larger than $3\times 10^{-11}$~G~\cite{dolag2004} outside large clusters,  the deflections of  iron nuclei become important  for energies below $1\times 10^{19}$~eV.  Therefore we only consider nucleons below this energy.

 When $E_{\rm cut} > 2.5 \times 10^{18}$ eV, besides fitting the spectrum above $E_{\rm cut}$,   we require that the spectrum we predict is never
above the measured spectrum at energies   between $2.5 \times 10^{18}$ eV and  $E_{\rm cut}$.

The plan of the paper is the following. In Section II,  we explain how
 we model the sources and the propagation of particles.
In Section III, we show the goodness of fit of the many models we consider.
 In Section IV  we show the average composition and the spectra of some of the models.
 We conclude in Section V.
  
\section{Modeling of the sources and particle  propagation }

We use a numerical code originally described in Ref.~\cite{kks1999}  
to compute the flux of GZK
photons produced by a uniform distribution of sources emitting originally
only protons or nuclei. The code uses the kinematic equation approach and
calculates the propagation of nuclei, nucleons, stable leptons  and photons
 using the standard dominant processes.  This is the
same numerical code as in Ref.~\cite{Gelmini:2007jy}, where 
the latest version of the code is described in detail.

 UHE particles lose their energy in interactions with the electromagnetic
background, which consists of CMB,  radio, infra-red and optical (IRO) components,
as well as EGMF.  Protons are sensitive essentially to the CMB only, while for
UHE photons and nuclei the radio and IRO components are respectively
important,  besides the CMB.  Secondary photons are always 
subdominant and thus do not contribute significantly to the fits. 
Therefore the radio background assumed is not important. 
 For the IRO background component we used the model
of Ref.~\cite{Stecker:2005qs}. This background is important 
for the photodisintegration of nuclei and to  transport the energy
 of secondary photons  in the cascade process 
from the  0.1 - 100~TeV energy range  to the  0.1-100~GeV energy range 
observed by EGRET,  and 
the resulting flux in this energy range is not sensitive to details
of the IRO background models.  
The possible deflection due to extragalactic magnetic fields is not included in the calculations.
These deflections could considerably extend the path of heavy nuclei below $1\times 10^{19}$~eV, but we do not consider the propagation of nuclei at these energies.

Notice that  if neutrons  are produced at the sources, the results at
 high energies are  very close to those obtained with protons. 
 The interactions of neutrons and protons with the intervening
backgrounds are almost identical and when  a neutron decays practically all
 of its energy goes to the final  proton (while the electron and neutrino are
produced with energies 10$^{17}$~eV or lower).

As is usual, we take the spectrum of an individual UHECR source  to
be of  the form:
\be  F(E) = f E^{-\alpha} ~~\Theta (Z E_{\rm max} -E)~,
 \label{proton_flux}
\ee
where $f$ provides the flux normalization, $\alpha$ is the spectral  index and
$E_{\rm max}$  ($Z E_{\rm max}$) is the maximum energy to which protons (or nuclei with charge $Z$) can be accelerated at the source.

We are implicitly assuming that the sources are astrophysical, since these are the
only ones which could produce solely protons (or neutrons) and nuclei as UHECR
primaries. Astrophysical acceleration mechanisms often result in $\alpha
\gsim 2$~\cite{AS2}, however, harder spectra, $\alpha \lesssim 1.5$ are also
possible, see e.g. Ref.~\cite{AS1.5}. In reality, the spectrum may differ from a power-law,
it may even have  a peak at high energies~\cite{peaks}. 
AGN cores could  accelerate protons with induced
electric fields, similar to what happens in a
linear accelerator, and this mechanism would produce  an 
almost monoenergetic proton flux, with energies as high as $10^{20}$~eV or
higher~\cite{mono}. Here, we consider the
power law index to be in the  range  $1 \le \alpha \le 2.7$.
An injected proton spectrum with $\alpha \geq 2.5$ does not require an extra contribution
to fit the UHECR data, except at very low energies 
$E<10^{18}$ eV~\cite{Berezinsky:2002vt}.
For $\alpha \le 2$ an extra low energy component
(LEC) is required to fit the UHECR data at $E<1 \times10^{19}$~eV.
Here we will consider values of $E_{\rm max}$ up to $10^{21}$~eV.

 We assume a standard cosmological model with  a Hubble constant
$H=70$~km~s$^{-1}$~Mpc$^{-1}$, a dark energy density (in units of the
critical density) $\Omega_{\Lambda}= 0.7$ and a dark matter density
$\Omega_{\rm m}=0.3$. The total source density in this model can be
defined by
\be n(z) = n_0 (1+z)^{3+m}~  \Theta (z_{\max}-z) \Theta (z-z_{\min}) \,,
\label{sources}
\ee
where $m$  parameterizes the source density evolution, in such a   way
that $m=0$ corresponds to non-evolving sources with constant density
per comoving volume,  and $z_{\min}$ and $z_{\max}$ are respectively
the redshifts of the closest and most distant sources. 

 The energy of the background photons
increases linearly with $(z+1)$ thus the GZK energy, about  $3\times 10^{19}$ eV at $z=0$, decreases
 as $1/(z+1)$ at redshift $z$. Moreover, the particles produced with that
energy at redshift z will arrive to us with energy redshifted as $1/(z+1)$, 
namely with characteristic energy $E = 3 \times 10^{19}$ eV$/(z+1)^2$.
This means that for $z>1$, $E<(3/4) \times 10^{19}$ eV, and for
 $z>2$, $E<(3/9) \times10^{19}$ eV.
 We conclude that sources with $z>1 $ have a negligible contribution  to the UHECR flux
above $1\times 10^{19}$~eV and those with $z>2$ do not contribute above $3\times 10^{18}$~eV. 
Thus any value of $z_{\max} \geq$ 1 or 2, respectively would give  the same results.

 We have considered several possible values of $m$, i.e.
$m=4, 2, 0, -2$ in this paper.  The fast and slow star formation rate
evolution models of Ref.~\cite{Stecker:2005qs} have $m=4$  and  $m=3$
respectively at  $z<1$  (and become constant close to  $z=1$ up to  $z>5$). 
The evolution of radio galaxies and
AGNs~\cite{dp90}, is somewhat faster than $m=3$ below $z=2$ 
(reaches a maximum at a $z$ between 2 and 3 and then
 decreases-see Fig.~6 of~\cite{De Marco:2005kt}). Smaller  positive
values of $m$ up to $m=0$, correspond to an older star population
evolution and is taken here as a lower limit to the value of $m$ at
low redshifts for protons. Negative values of $m$ have been mentioned
in the literature only for very massive clusters, which only formed
recently. However, accretion shocks in clusters might accelerate heavy
nuclei but not protons to the energies necessary to account for the
ultrahigh energy cosmic rays~\cite{Inoue:2007kn}.

The value of $z_{\min}$ is connected to the
density of sources. Quite often in the literature the minimal distance to 
the sources is assumed to
be negligible (i.e. comparable to the interaction length).  We  also consider 
non-zero minimum distances of up to 50 Mpc  ($z_{\rm
min}= 0.01$), as inferred from the
small-scale clustering of events seen in the AGASA data~\cite{AGASA_clusters}. 
Contrary to AGASA, HiRes does not see a clustering component in its own
data~\cite{HiRes_clusters}. The combined dataset shows that clustering still
exists, but it is not as significant as in the data of AGASA
alone~\cite{agasa_hires}.  Note, that the non-observation of clustering in the
HiRes stereo data does not contradict the result of AGASA, because of the
small number of events in the sample~\cite{agasa_hires_ok}.
Assuming proton primaries and a small EGMF (following Ref.~\cite{dolag2004}),
it is possible to infer the density of the 
sources~\cite{agasa_hires_ok, sources} 
from the clustering component of UHECR. AGASA data alone
suggest a source density of $2\times 10^{-5}$~Mpc$^{-3}$, which makes
plausible the existence of one source within 25 Mpc of us. However, the HiRes
negative result on clustering requires a larger density of sources and, as a
result, a smaller distance to the nearest one of them. Larger 
values of the EGMF
(as found in Ref.~\cite{Sigl:2004yk}), and/or some fraction of iron
in the UHECR, have the effect of reducing the required number of sources and,
consequently, increasing the expected distance to the nearest one.

Most of the energy in GZK photons cascades down to below the pair production 
threshold for photons on the CMB and infrared backgrounds.  In general, for $\alpha<2$ 
the diffuse extragalactic gamma-ray flux measured by EGRET~\cite{EGRET}
at GeV energies may impose a constraint on the GZK photon flux at high energies, 
which we  take into account and found not relevant for any of the models we study here.

\section{Goodness of fit of different source models}

In this section we estimate the flux predicted by the models by fitting the
Pierre Auger UHECR spectrum.  We proceed using the method explained in Ref.~\cite{Gelmini:2007sf}.

We  fit the Pierre Auger UHECR data assuming many different injected spectra. 
We assume an injected spectrum given by Eq.~\ref{proton_flux},
a uniform distribution of sources with a density as in Eq.~\ref{sources} with
$z_{\rm max} = 3$ and,   $z_{\rm min} = 0$ or 0.005 or 0.01 and $m= 4$ or  2 or 0 or -2.
We consider then many different spectra resulting from changing   the slope
$\alpha$ and the maximum energy $E_{\max}$ in 
Eq.~\ref{proton_flux} within the ranges 
$1 \leq \alpha\leq 2.7$ and $10^{19} {\rm eV}\leq  E_{\max}\leq  1.28 \times 10^{21}$~eV  
in steps $\alpha_n=1+0.1 n$, with $n=0$ to 17 and 
$E_{\ell}=1 \times 10^{19}eV \times 2^\ell$, with
$\ell=0$~to~7. 
For each one of the models so obtained we compute the predicted UHECR spectrum
arriving to us from all sources.

In order to compare  the predicted flux with the data, we  also take
into account the experimental error in the energy determination 
as proposed in Ref.~\cite{Albuquerque:2005nm}. We take a
lognormal distribution for the error in the energy reconstructed by the
experiment with respect to the true value
of energy of the UHECR coming into the atmosphere. To find the
expected flux we convolute the spectrum predicted by each model with
the  lognormal distribution in energy with 
the width given by the Pierre Auger energy error $\Delta E/E =
8$\%~\cite{ICRC2007-auger-energy-determination} (the parameter
$\sigma$ in Eq. (5) of  
Ref.~\cite{Albuquerque:2005nm}, the standard deviation of log$_{10} E$, is
$\sigma = (\Delta E/E)/$ln(8)$ \simeq (\Delta E / E)/ 2.08$).
This procedure results in small but non-negligible changes in the
predicted spectra which are then compared to the  observed spectrum. In particular, there are 
events predicted with an energy larger than the maximum  injected energy $Z E_{\max}$. Somewhat arbitrarily
we consider the energy beyond which no event is predicted to be $(1 + 10 \Delta E/E) Z E_{\max}$.
Moreover, we take into account that there is about a factor of 2
between the energy of a photon event and the energy measured if the
event is reconstructed assuming it is a
proton~\cite{auger-sd-photon-limit}. Thus we divide the energy of the
predicted GZK photon energy by 2 before comparing it with the
observed Pierre Auger spectrum.  However, the GZK photons are always subdominant in the flux of
UHECR~\cite{Gelmini:2007jy, Gelmini:2007sf}
 thus  they do not affect the goodness of the fits (and at present the GZK photon fractions are not constrained by Auger upper bounds- see Fig. 18 of Ref.~\cite{Gelmini:2005wu}).
 
\begin{figure}
\includegraphics[width=0.325\textwidth,clip=true,angle=0]{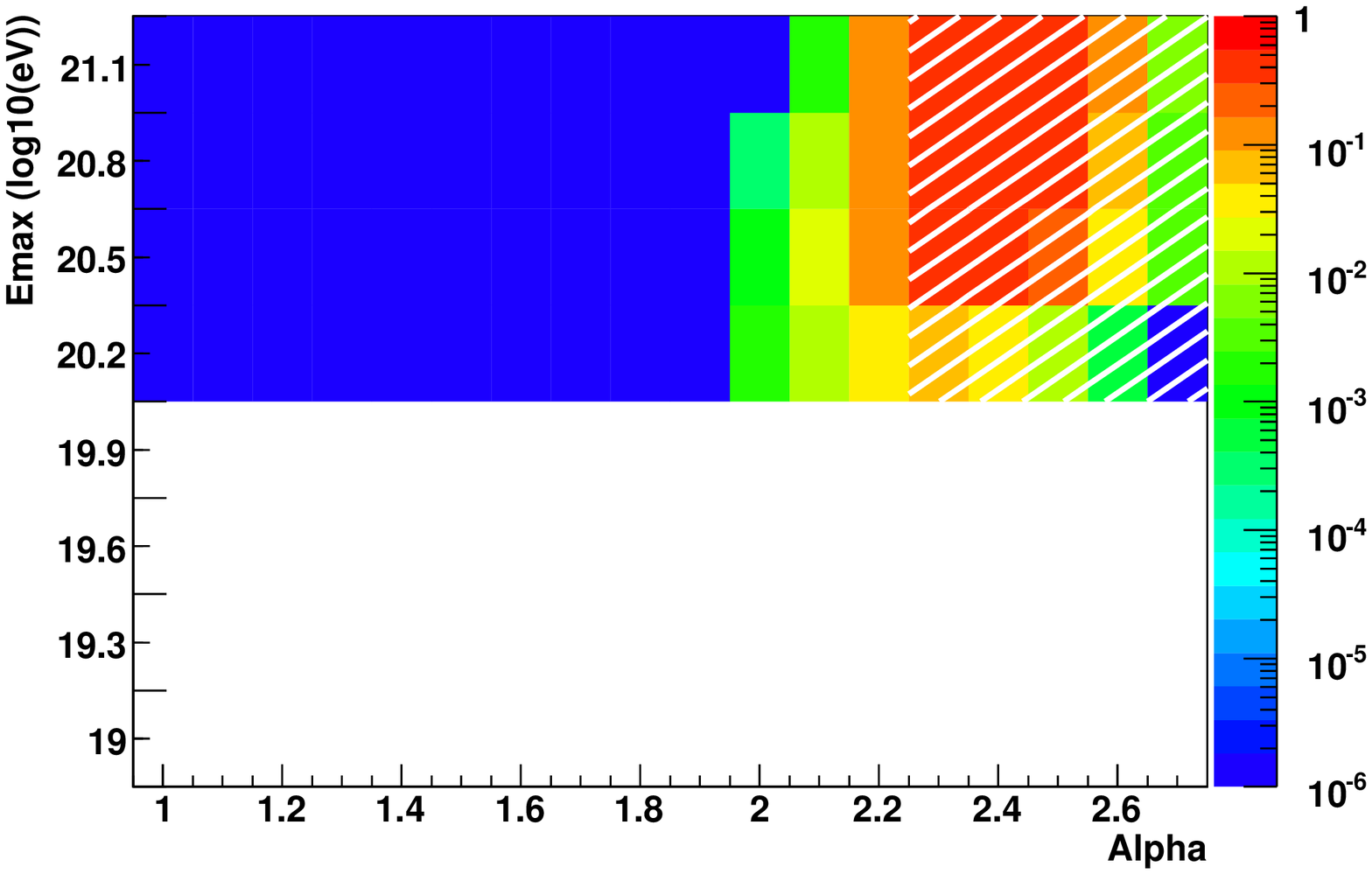} 
\includegraphics[width=0.325\textwidth,clip=true,angle=0]{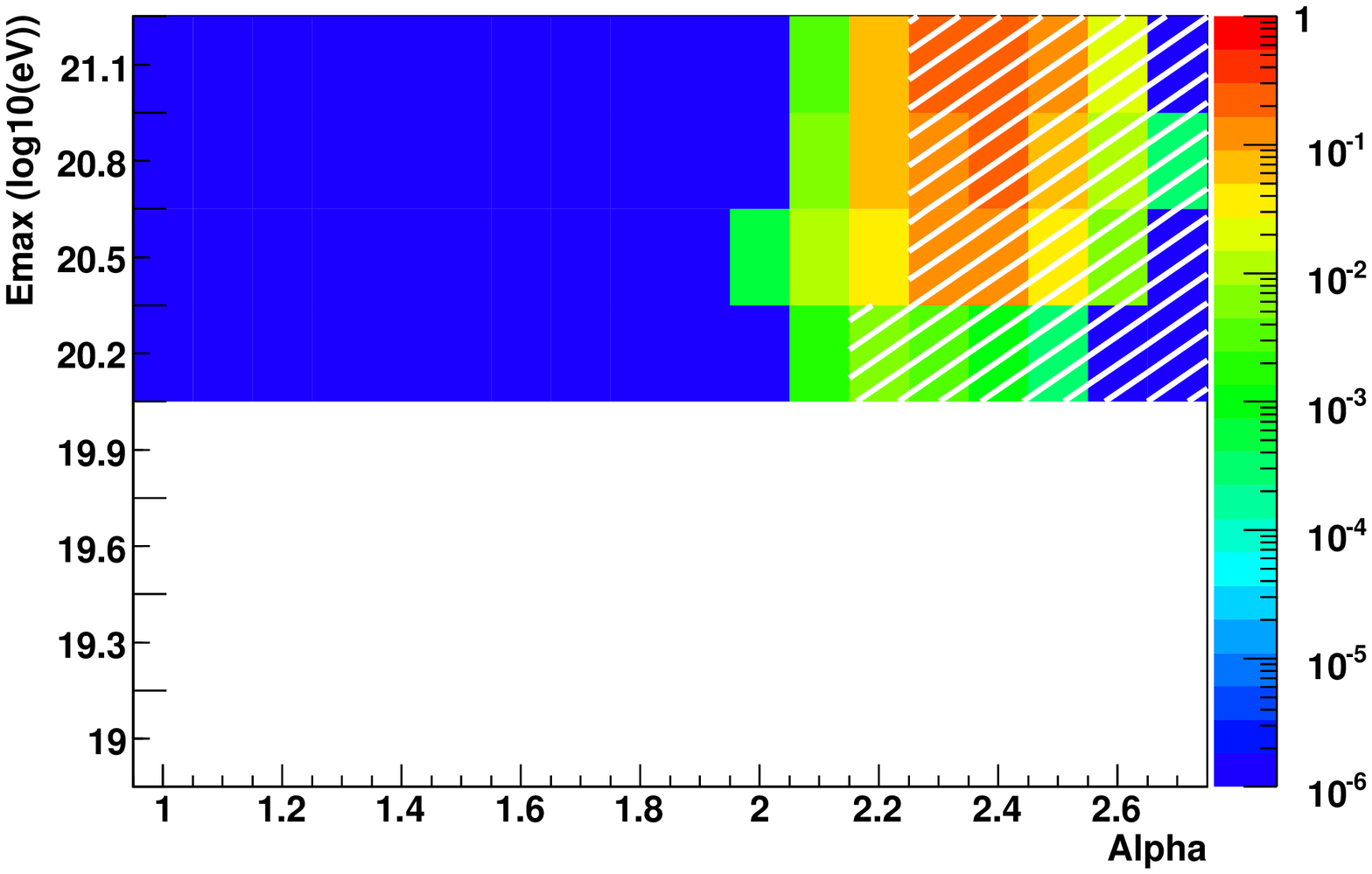} 
\includegraphics[width=0.325\textwidth,clip=true,angle=0]{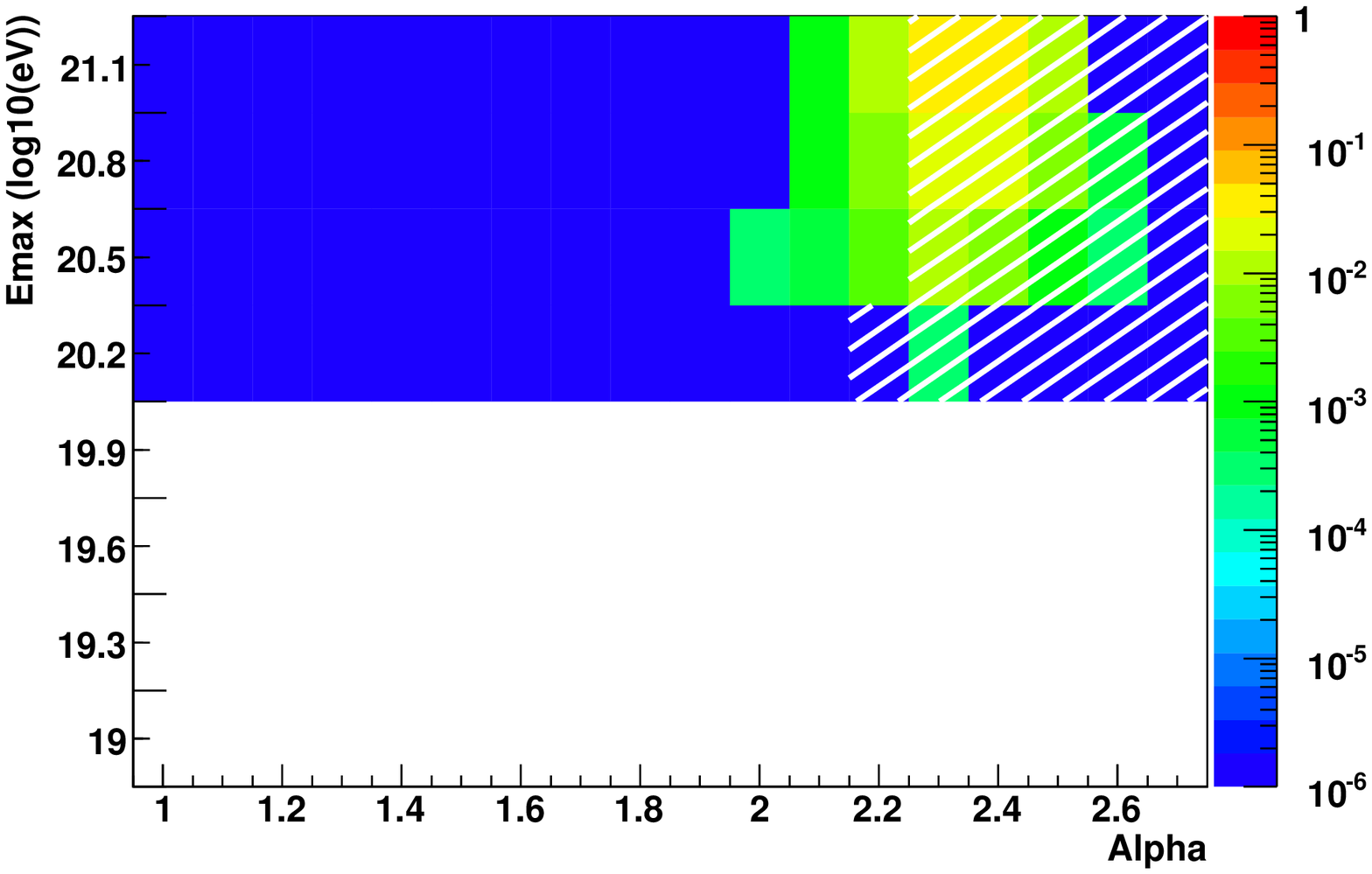}
\includegraphics[width=0.325\textwidth,clip=true,angle=0]{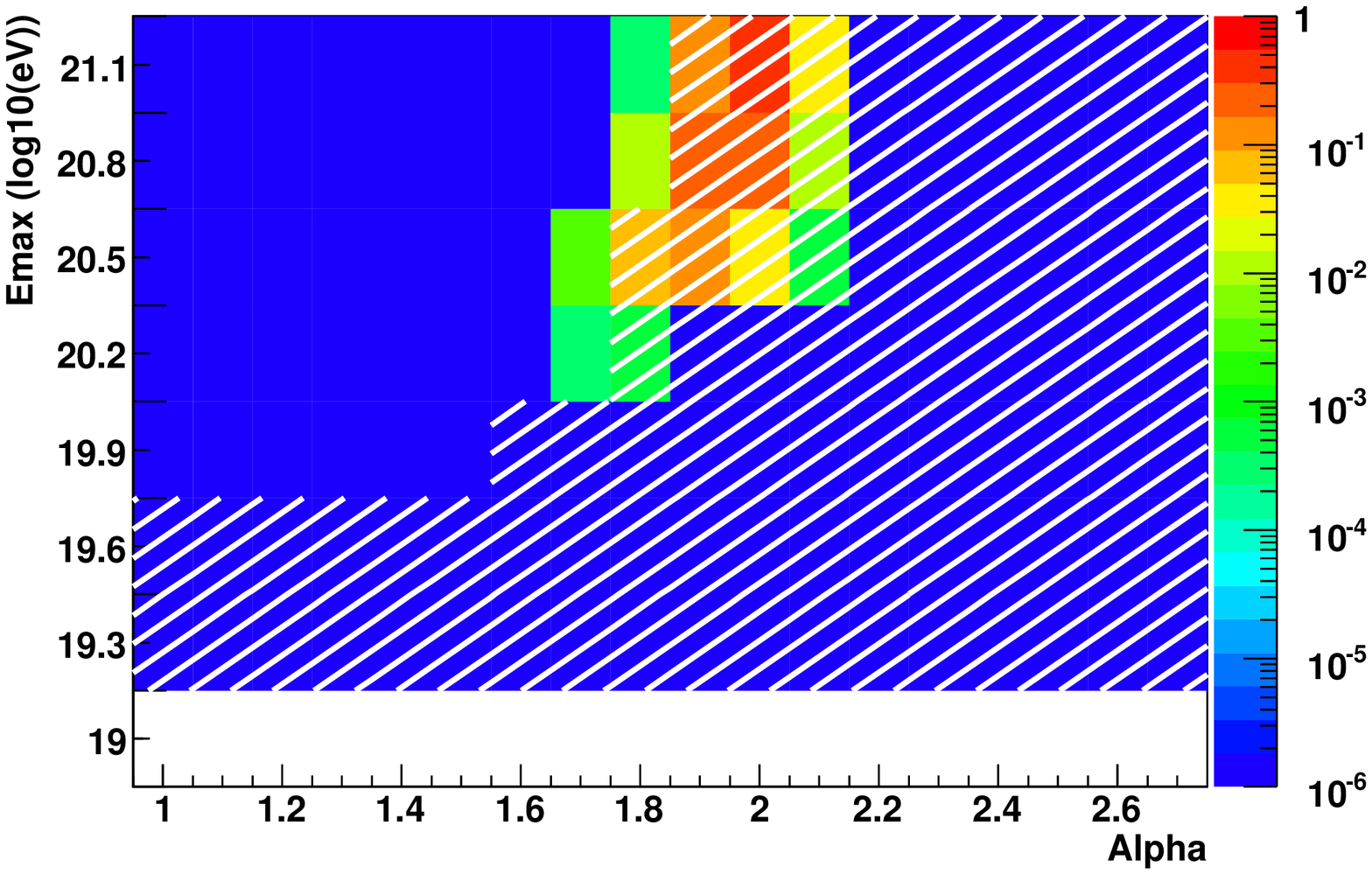} 
\includegraphics[width=0.325\textwidth,clip=true,angle=0]{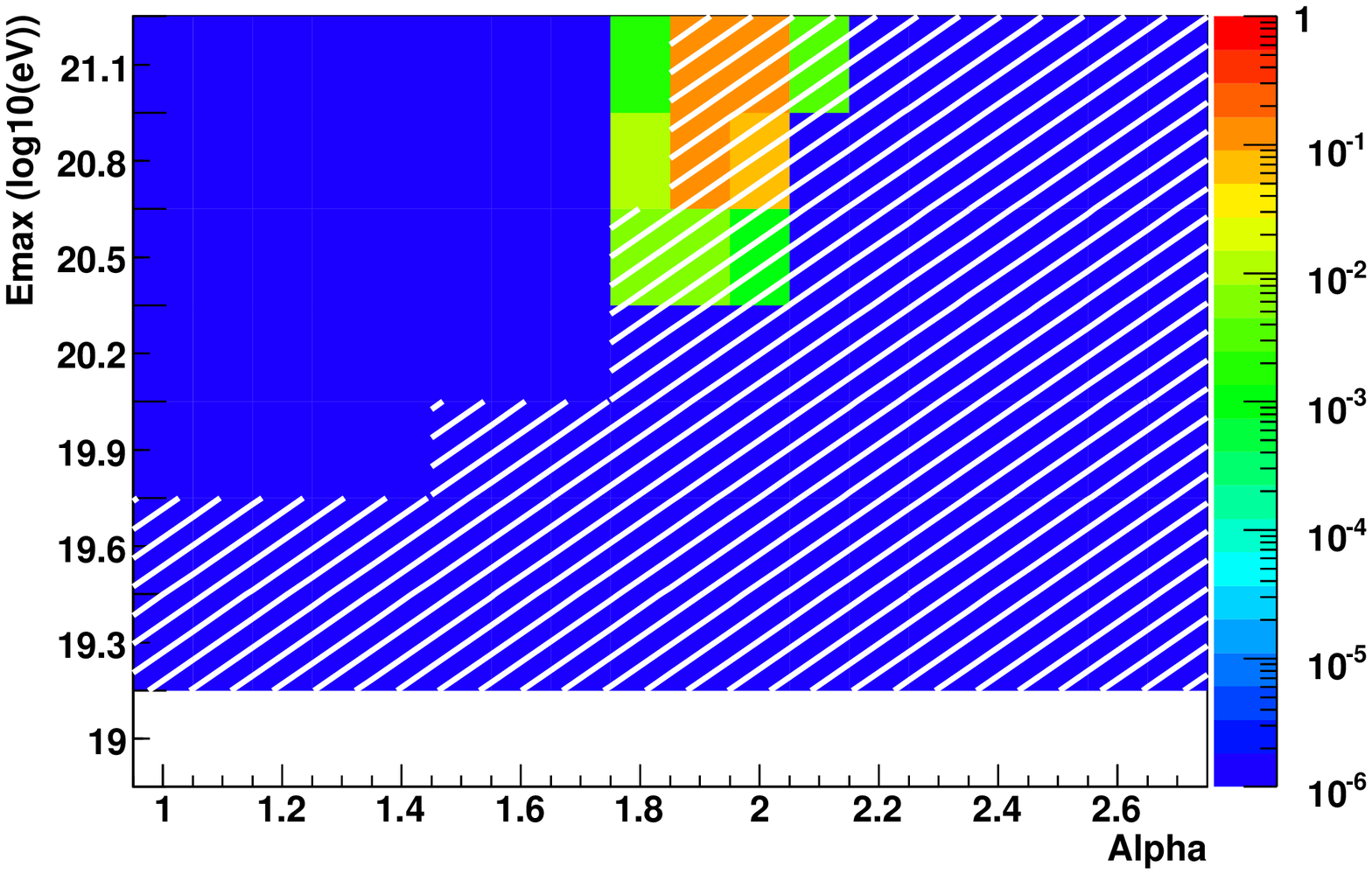} 
\includegraphics[width=0.325\textwidth,clip=true,angle=0]{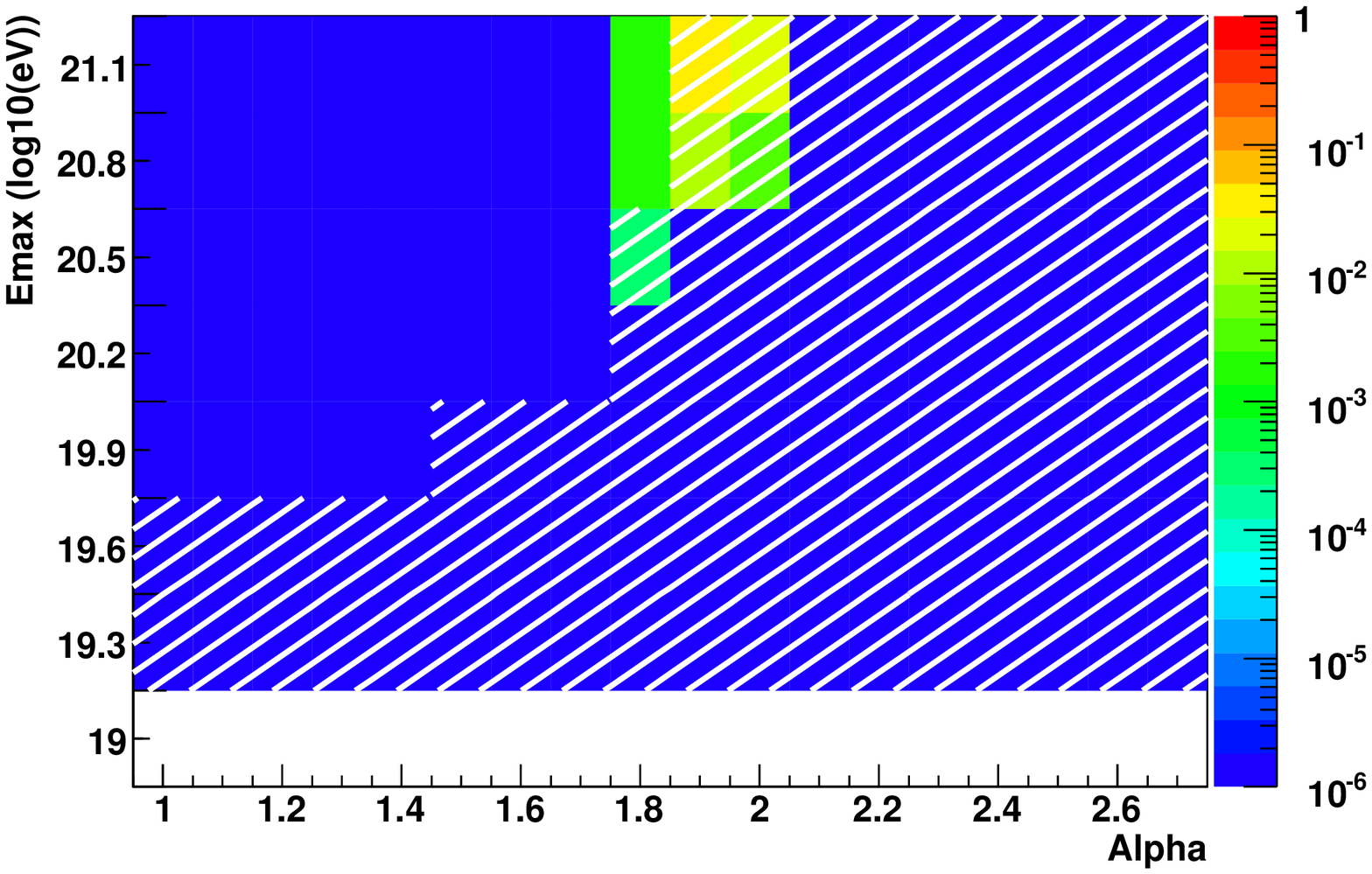} 
\includegraphics[width=0.325\textwidth,clip=true,angle=0]{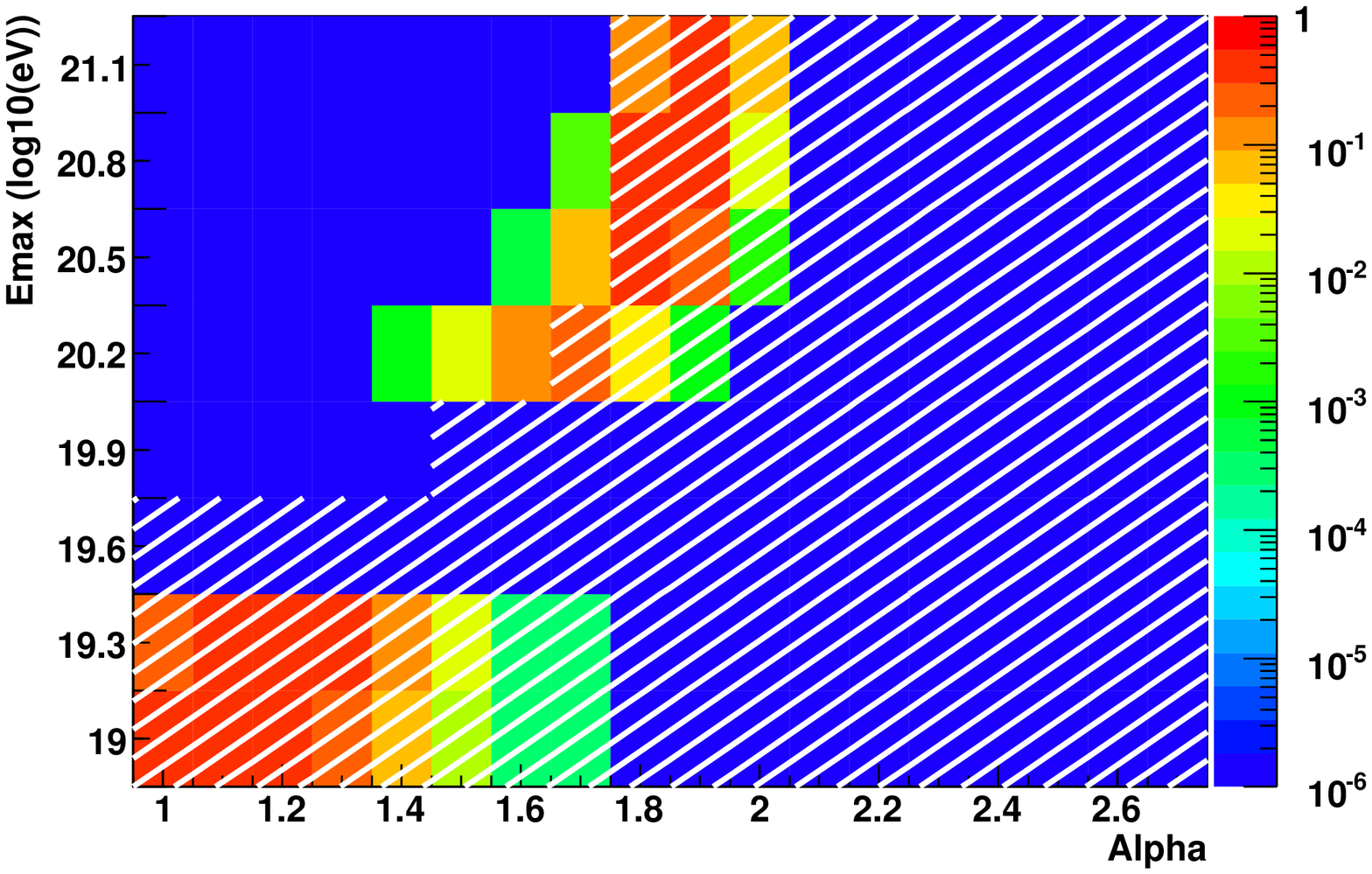} 
\includegraphics[width=0.325\textwidth,clip=true,angle=0]{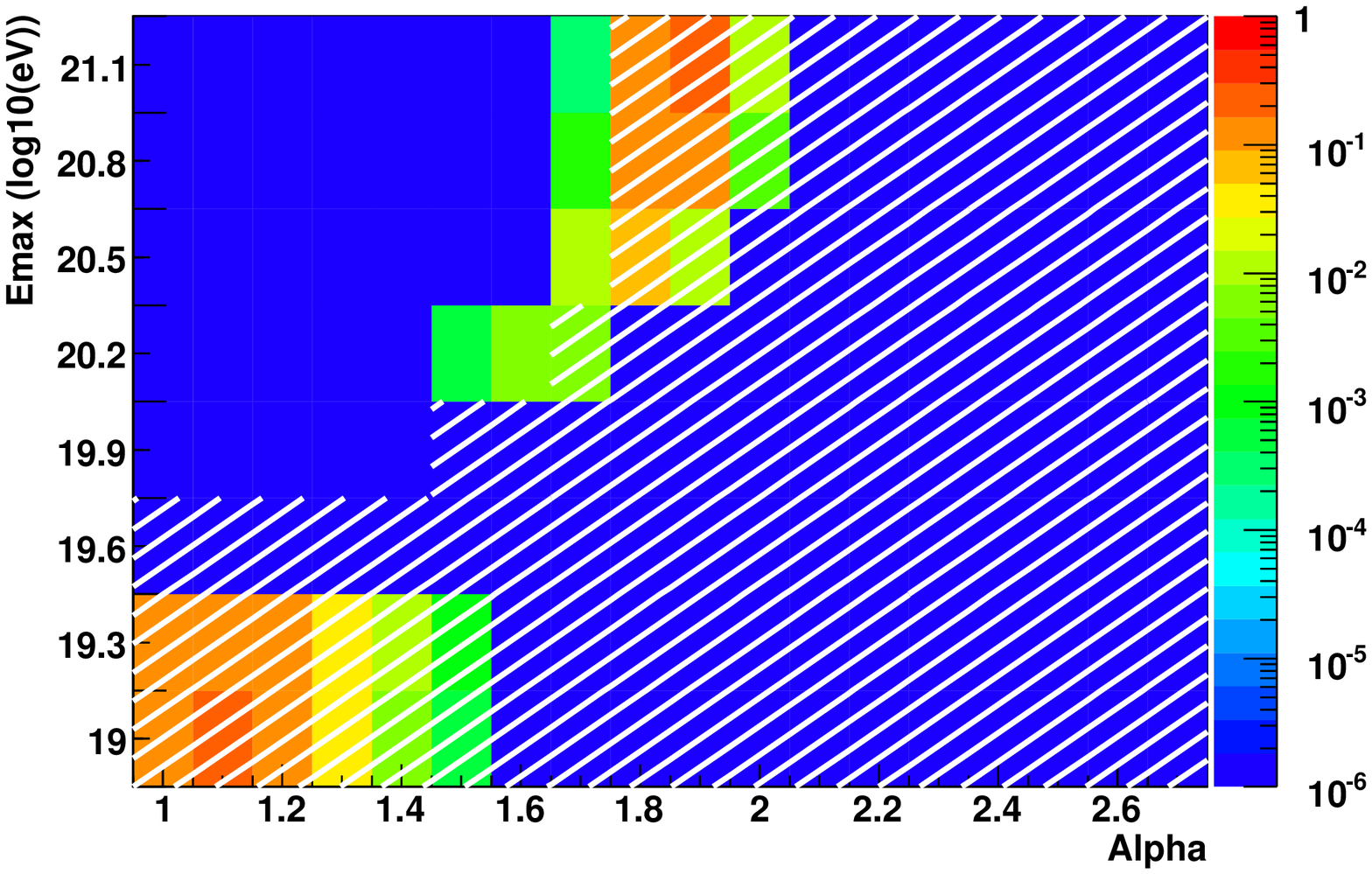} 
\includegraphics[width=0.325\textwidth,clip=true,angle=0]{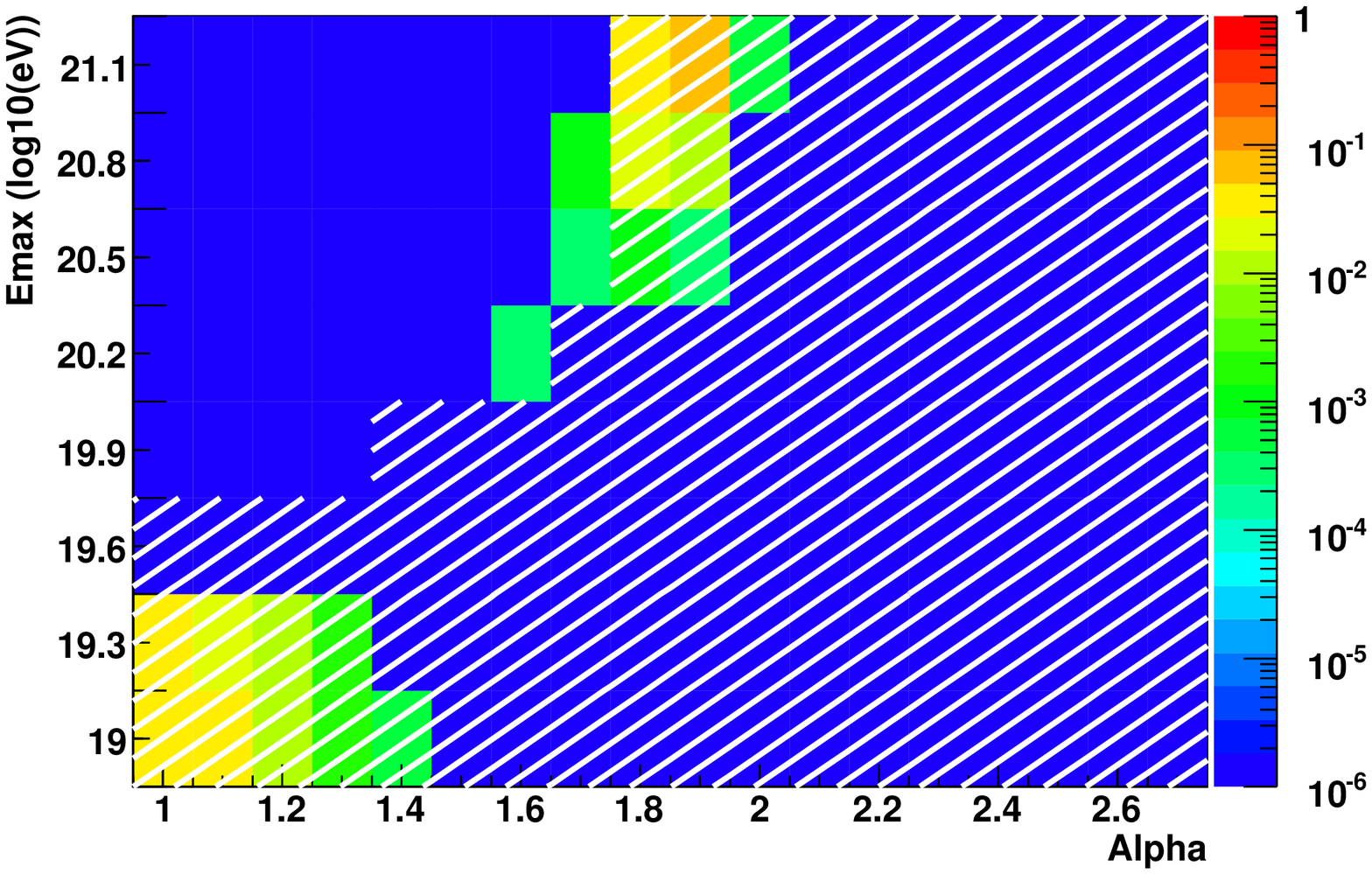} 
\caption{Color coded  $p$-value plots as function of $E_{\rm max}$ and 
  $\alpha$ for $E_{\rm cut}= 1 \times 10^{19}$~eV and $m=4$, for p, O, or Fe emitted at the sources
  (top to bottom) and  $z_{\rm min}=$ 
  0, 0.005, 0.01 (left to right).  White regions for p  are eliminated because of energetic reasons.
Cross hatched regions eliminated by the requirement at $2.5 \times 10^{18}$ eV $<E< E_{\rm cut}$ (see text). Only orange and red regions have $p >0.05$.}
\label{p-values-filter-Emax-alpha}
\end{figure}

\begin{figure}
\includegraphics[width=0.325\textwidth,clip=true,angle=0]{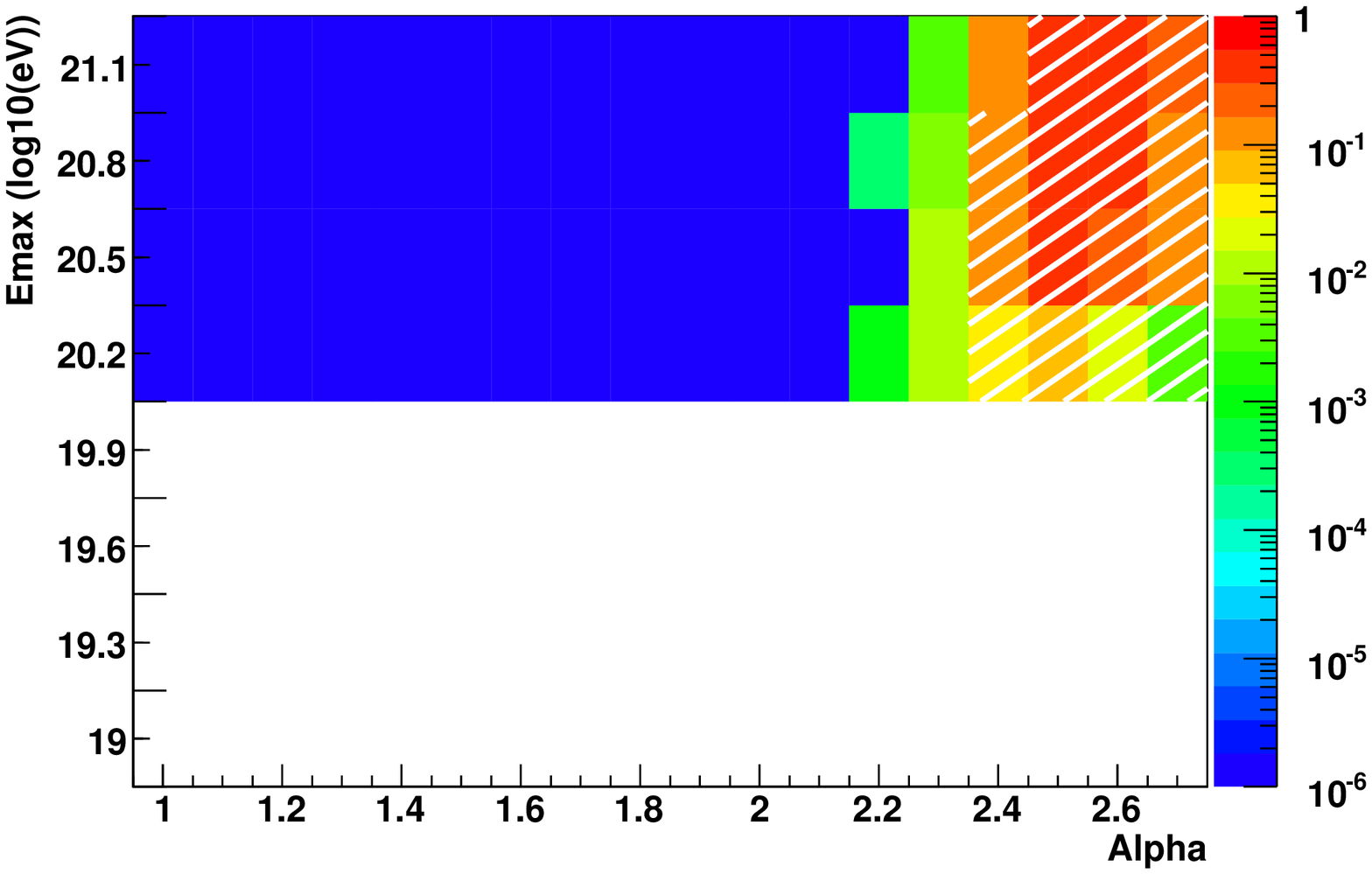} 
\includegraphics[width=0.325\textwidth,clip=true,angle=0]{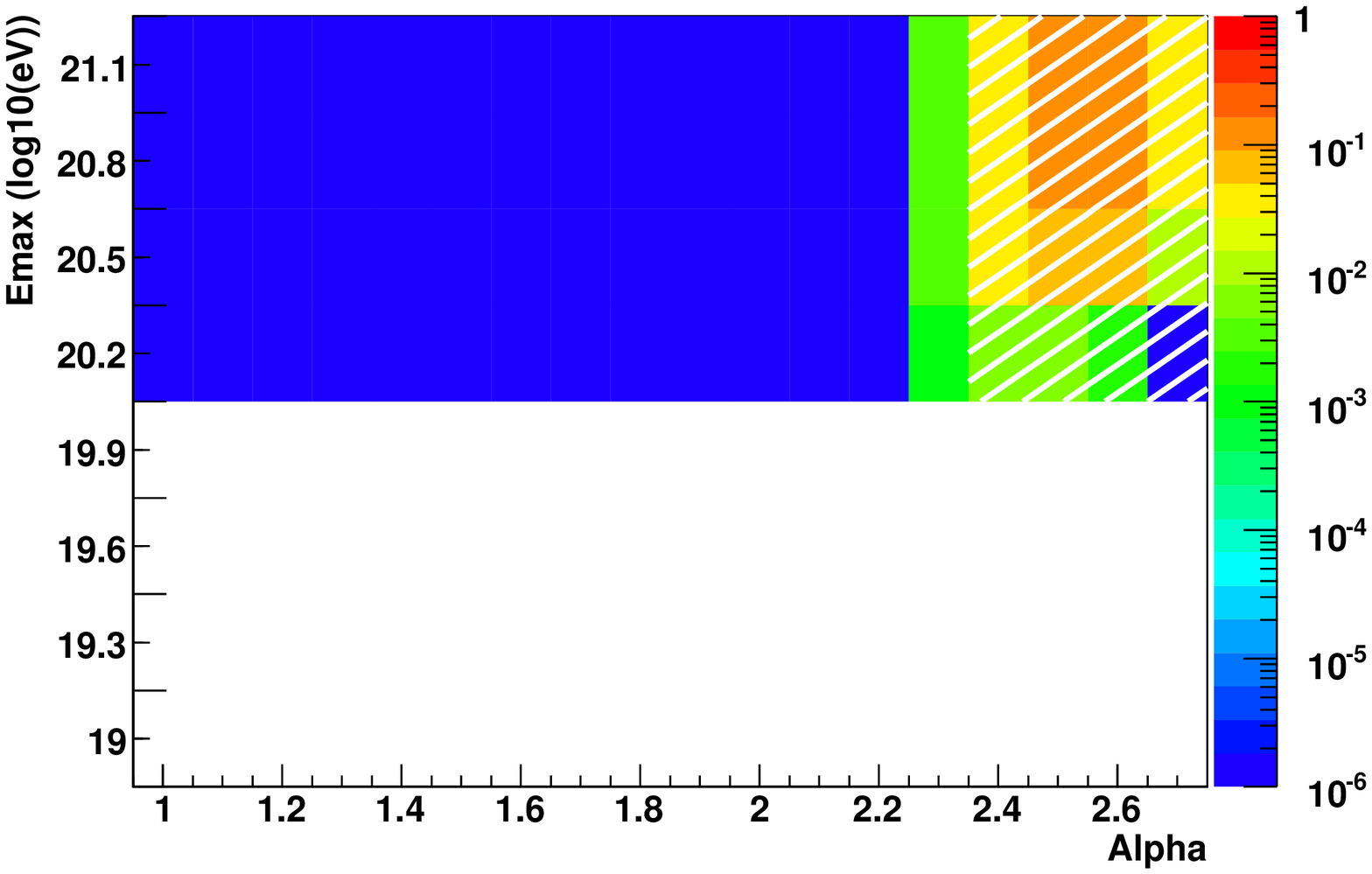} 
\includegraphics[width=0.325\textwidth,clip=true,angle=0]{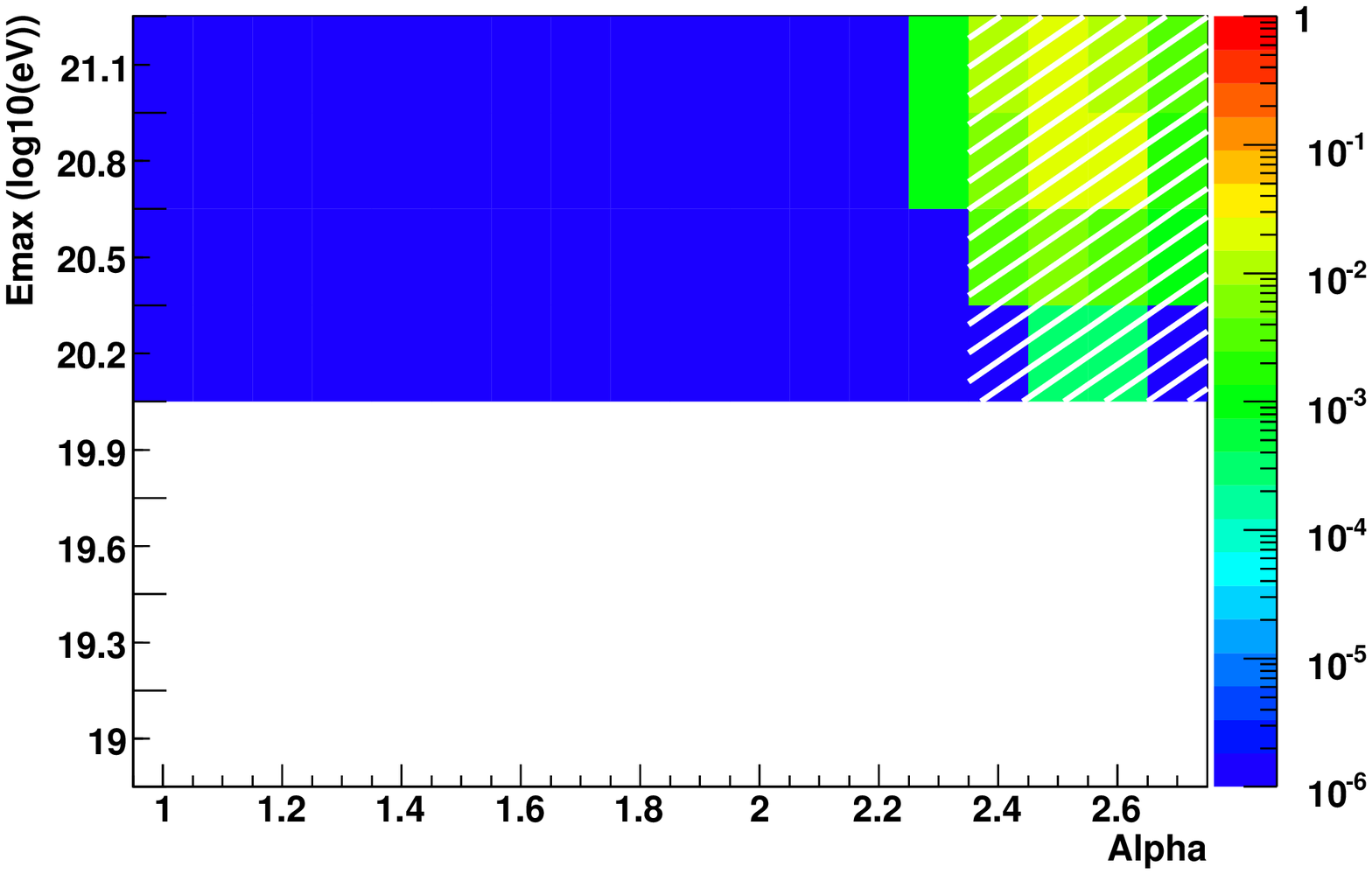} 
\includegraphics[width=0.325\textwidth,clip=true,angle=0]{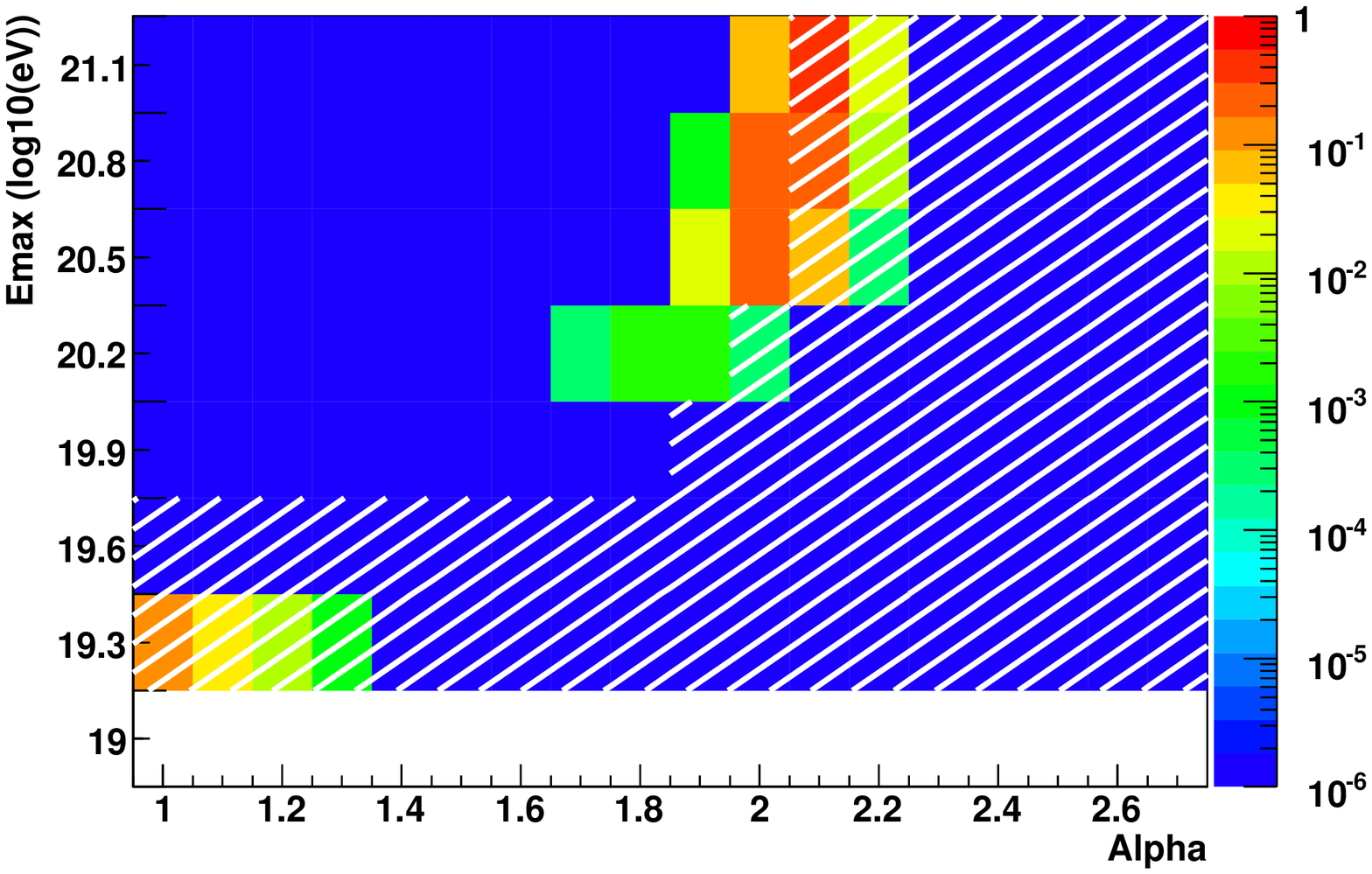} 
\includegraphics[width=0.325\textwidth,clip=true,angle=0]{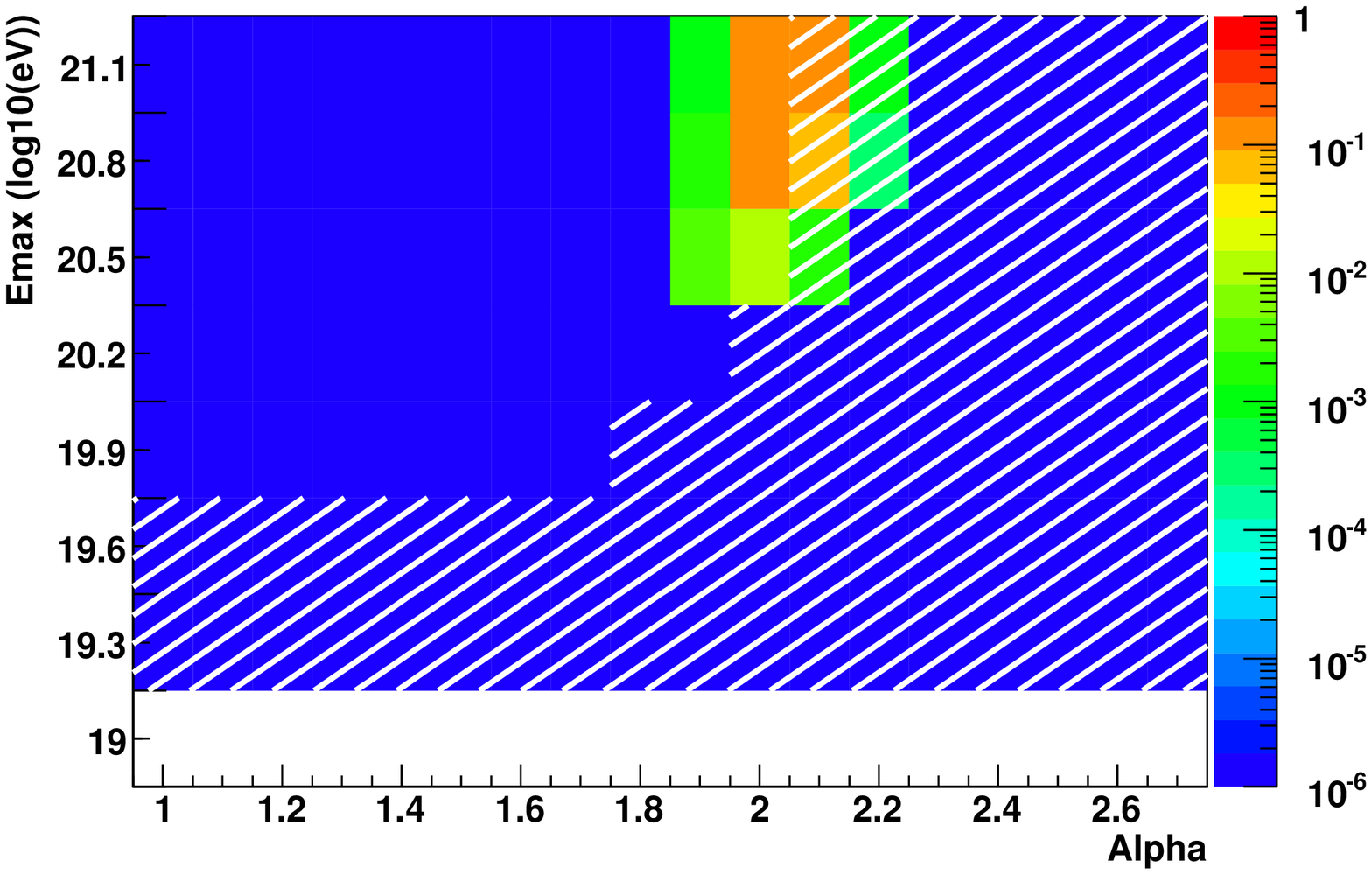} 
\includegraphics[width=0.325\textwidth,clip=true,angle=0]{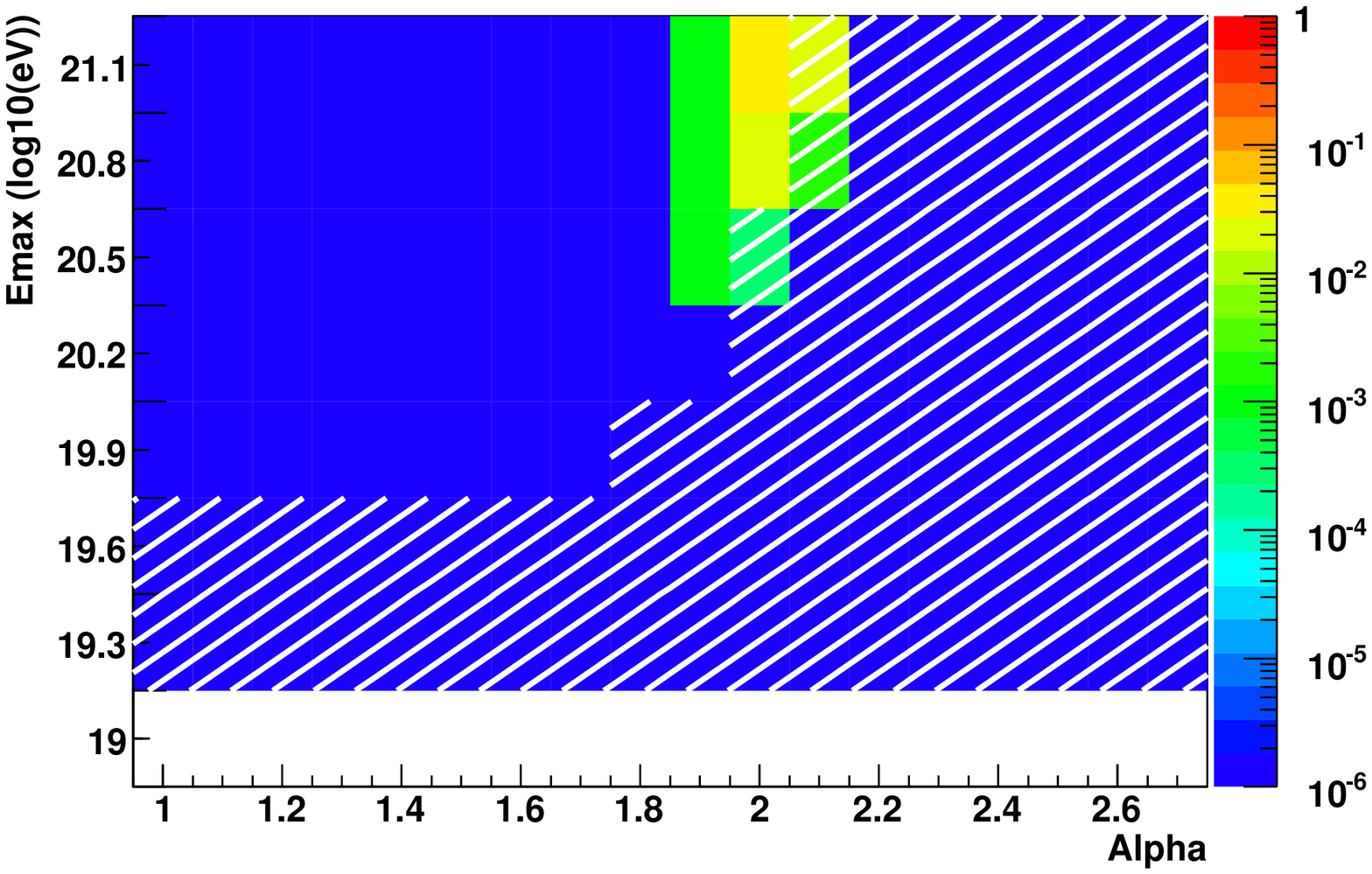}
\includegraphics[width=0.325\textwidth,clip=true,angle=0]{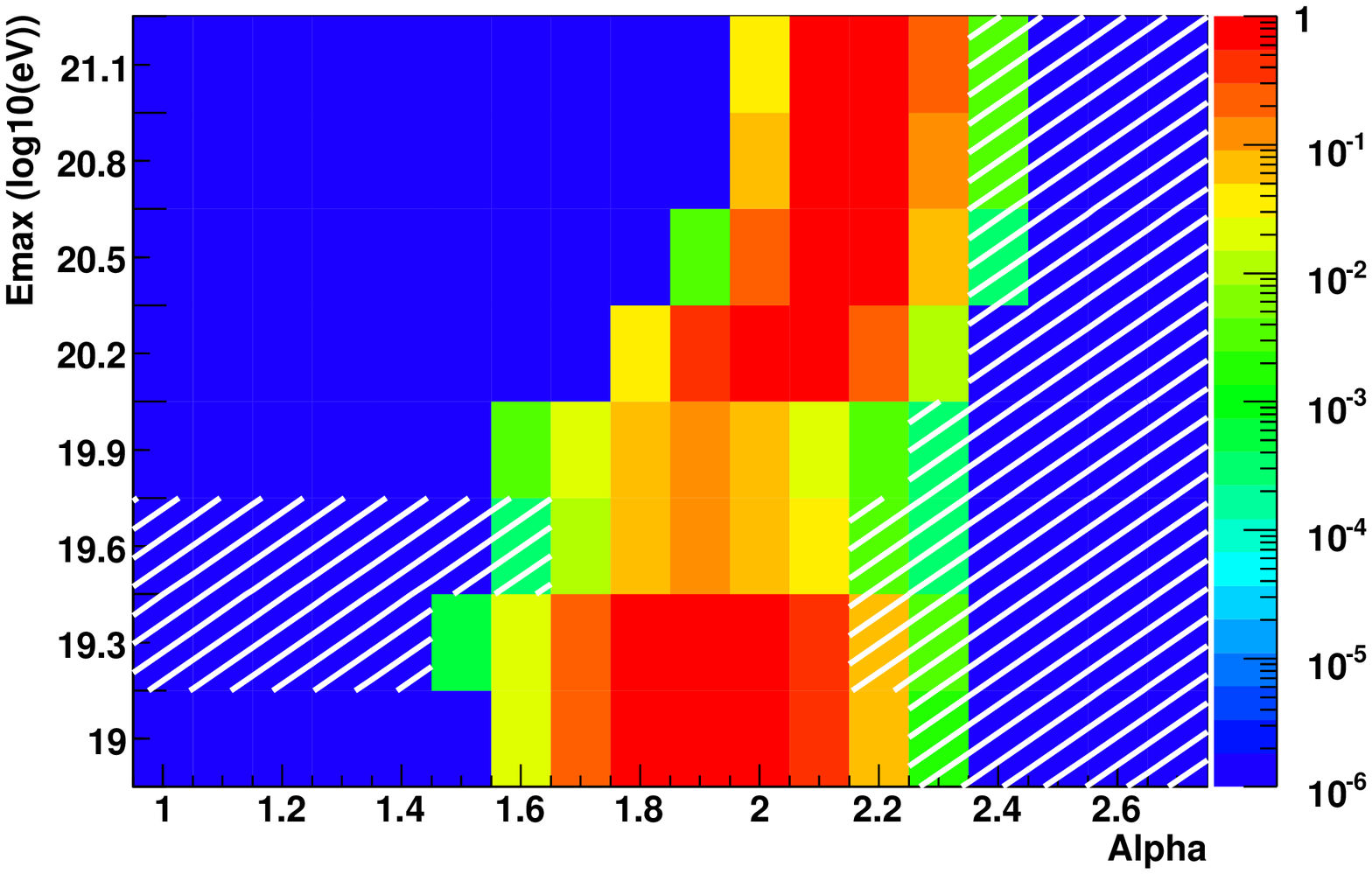} 
\includegraphics[width=0.325\textwidth,clip=true,angle=0]{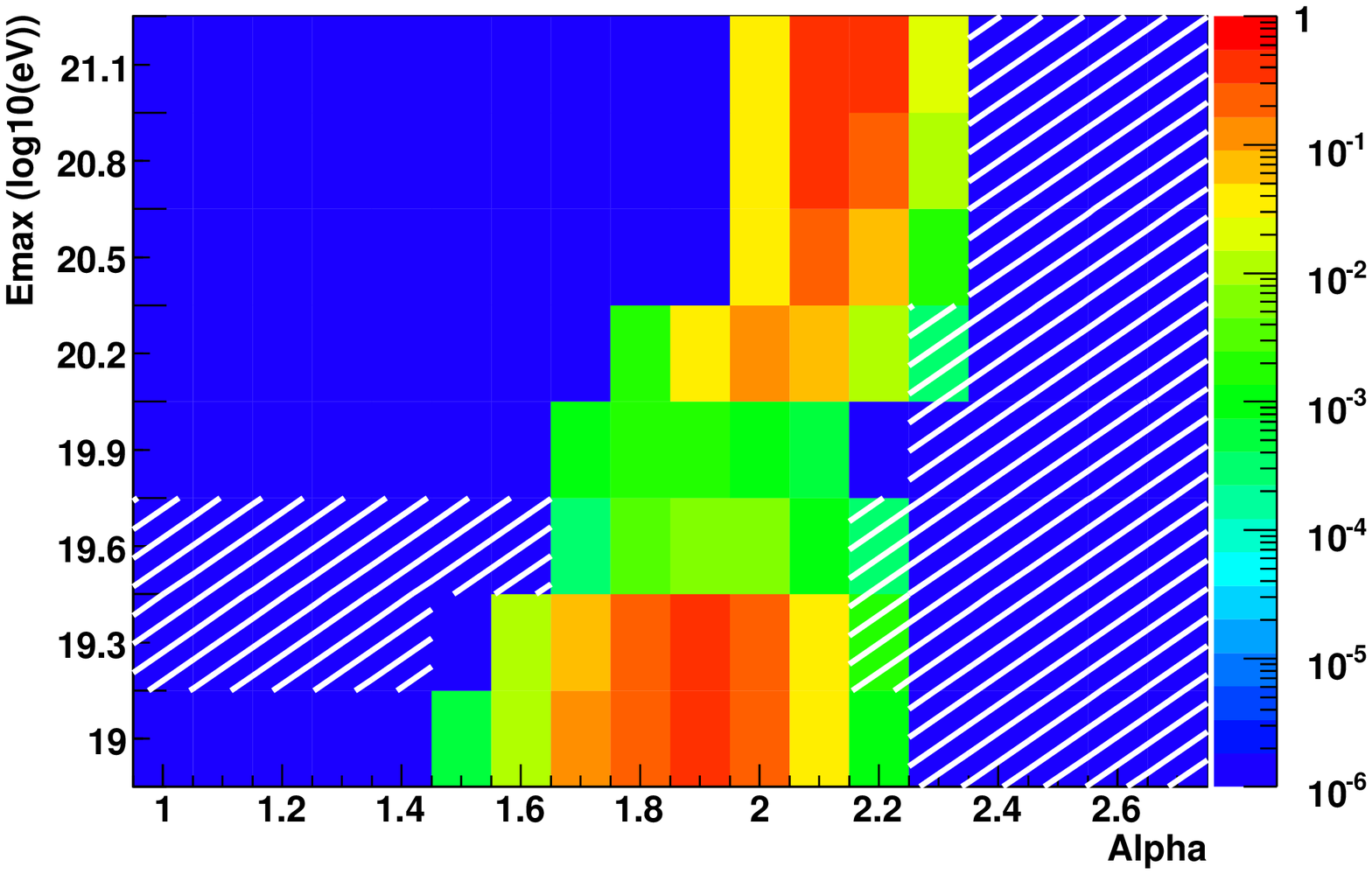}
\includegraphics[width=0.325\textwidth,clip=true,angle=0]{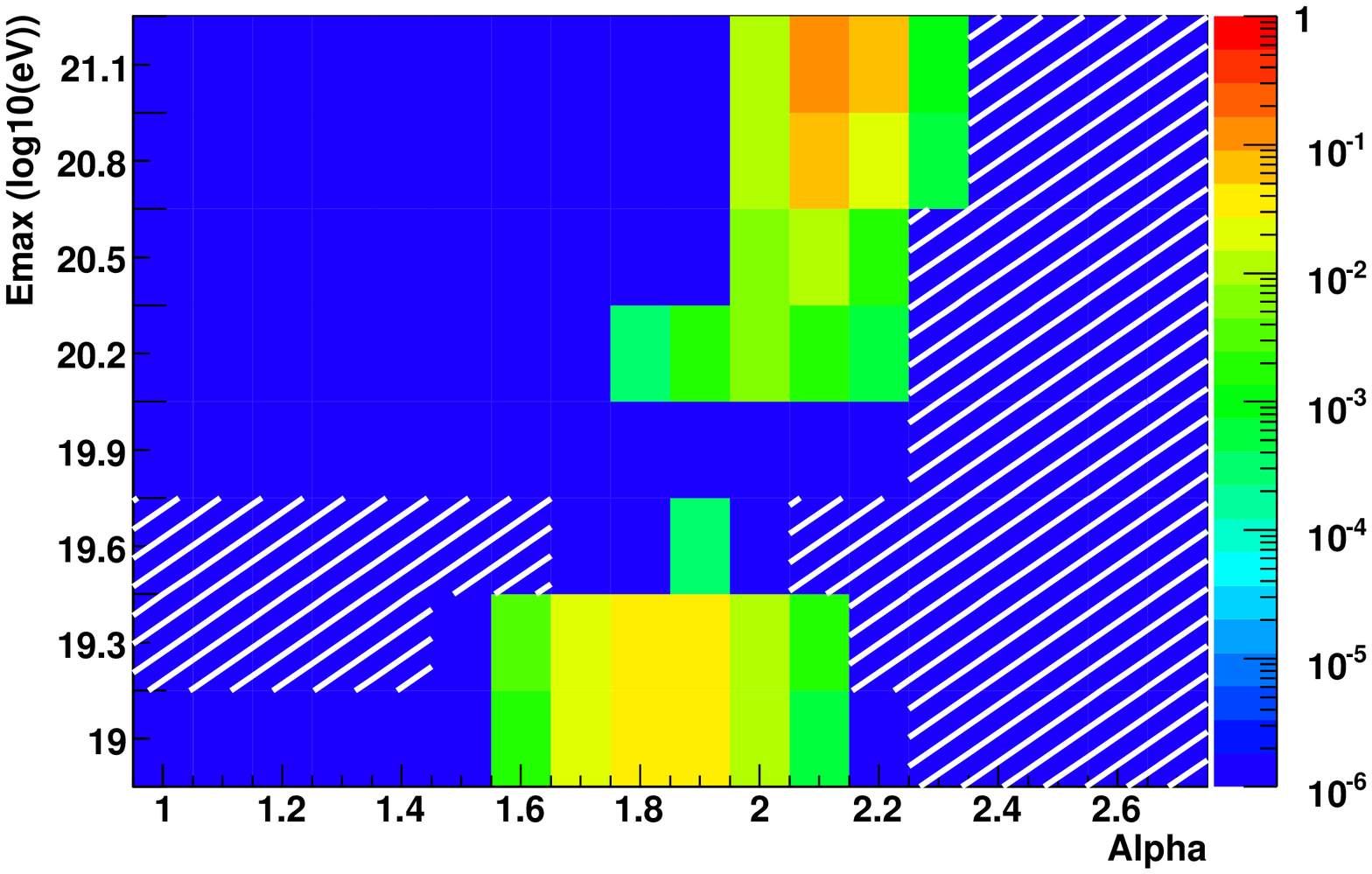} 
\caption{Same as Fig.~\ref{p-values-filter-Emax-alpha} but for
  $m=0$. Notice that the best fit regions have shifted to higher values of
  $\alpha$.}
\label{p-values-filter-Emax-alpha-mzero}
\end{figure}

With each predicted spectrum we fit the UHECR data from
 $E_{\rm cut}$ up to a bin past the last published bin of the
spectrum (which is the $10^{20.3}$~eV bin of the Pierre Auger
Observatory).  The extra bin extends from the maximum experimental
point of the observed spectrum,
$10^{20.4}$~eV~\cite{ICRC2007-auger-energy-spectrum} (which is also
empty) to $(1 + 10 \Delta E/E) Z E_{\max}$ (where $Z E_{\max}$ is the maximum
energy assumed for the injected spectrum in Eq.~\ref{proton_flux}).
We do this because the assumed injection spectrum could produce an
event in this bin even though the experiment did not observe one.  If
the maximum possible energy, $(1 + 10 \Delta E/ E) Z E_{\max}$, is
less than the maximum bin of the published spectrum the additional
bin is not needed and therefore not added.  In certain assumed
injection spectra the maximum possible energy is less than the energy
of the most energetic event observed.  In this case the assumption is
not valid on the face of it and therefore immediately disqualified.
Situations like this can be seen in
Fig.~\ref{p-values-filter-Emax-alpha}~and~Fig.~\ref{p-values-filter-Emax-alpha-mzero}
as the empty regions.

 We also change $E_{\rm cut}$ and fit the UHECR
data from $2.5\times 10^{18}$~eV (only with injected protons) or $4\times 10^{19}$~eV and compare
the results with those of $1\times 10^{19}$~eV.  We show how this
affects the goodness of the fit in Fig.~\ref{p-values-filter-Ecut}
using proton sources.

The expected number of events in each bin between $E_{\rm cut}$ and the
maximum energy bin is computed using the exposure of the
Pierre Auger Observatory,
5165~km$^2$~sr~yr~\cite{ICRC2007-auger-energy-spectrum}.  The aperture
remains constant with increasing energy.

Fitting the UHECR data with a predicted spectrum follows a
procedure similar to that of Ref.~\cite{Fodor-K-R} applied to the
bins just mentioned.
We compare the observed
number of events in each bin with the number of events predicted by
the models and choose the value of the
parameter $f$ in Eq.~\ref{proton_flux}, i.e. the amplitude of the
injected spectrum, by maximizing the Poisson likelihood function.
This is equivalent to minimizing $-2 \ln{\lambda}$, (i.e. the
negative of the log likelihood ratio)~\cite{statistics}. This
procedure amounts to choosing the value of $f$ so that the mean total
number of events predicted (i.e. the sum of the average predicted
number of events in all fitted bins) is equal to the total number
of events observed. We then compute, using a Monte Carlo technique, the
goodness of the fit, or $p$-value of the distribution, defined as the
mean fraction of hypothetical experiments (observed spectra) with the
same fixed total number of events which would result in a worse, i.e. smaller,
Poisson likelihood than the one obtained (in the
maximization procedure that fixed $f$). These hypothetical experiments
are chosen at random according to the multinomial distribution of the
model (with $f$ fixed as described).  We have checked that this
procedure when applied to bins with a large number of events gives the
same result as a Pearson's $\chi^2$ fit, both for the value of the
normalization parameter $f$ and for the goodness of fit.  A higher
$p$-value corresponds to a better fit since a greater number of hypothetical
experimental results would yield a fit worse than the one we
obtained.

We make one additional requirement on the fit to insure the
predicted flux does not exceed the observed flux at energies below
$E_{\rm cut}$  and above $2.5\times 10^{18}$ eV, the lowest energy of the published Auger spectrum. When $E_{\rm cut}> 2.5 \times 10^{18}$ eV, 
for each assumed spectrum (with $f$ fixed as
described above) we calculate the $\chi^2$ for the data at energies below
$E_{\rm cut}$  using only the data points in which the predicted flux
is above the observed flux (i.e. we take as zero the contribution to
the $\chi^2$ of each data point for which the predicted flux is below
the observed flux). We then require the $p$-value of the $\chi^2$ so
obtained to be larger than 0.05. This constraint eliminates many
combinations  of $\alpha$ and $E_{\rm max}$ values.  The regions eliminated by this
requirement are the cross hatched regions in
Fig.~\ref{p-values-filter-Emax-alpha}, 
\ref{p-values-filter-Emax-alpha-mzero}, \ref{p-values-filter-m-alpha} and \ref{p-values-filter-Ecut} .
This low energy constraint would, however, be too restrictive if somehow the extragalactic cosmic rays below some threshold energy between $2.5\times 10^{18}$ eV and  $E_{\rm cut}$ do not reach Earth --- for example, due to magnetic confinement at the source. In this case,  the deficit   of extragalactic flux below the threshold energy
 should  be made up by a (possibly galactic) LEC.

Fig.~\ref{p-values-filter-Emax-alpha} and
 \ref{p-values-filter-Emax-alpha-mzero} show in a logarithmic scale the
color coded $p$-value of the maximum Poisson likelihood value
obtained for each model as a function of $E_{\rm max}$ and $\alpha$,
for $m=4$ and $m=0$, respectively.  The top, middle and lower panels correspond to  proton,  oxygen, and  iron emitted by the sources, respectively, while the columns from left to right
 correspond to $z_{\min} =$~0, 0.005, 0.01,
respectively. Overall,  the cross hatched region  (in which the flux predicted
at energies $2.5\times 10^{18}$ eV $<E< E_{\rm cut}$ exceeds the observed one) includes many  regions of $E_{\rm max}$, $\alpha$ which would otherwise 
provide good fits (red and orange regions where  $p$-value $\ge0.05$).
In some instances the acceptable models lie just outside a cross hatched
region and in some others no acceptable models remain.  The blue, green-blue, or yellow regions do not provide good fits (if we choose
only $p$-value $\ge0.05$ to be acceptable). 

\begin{figure}
\hspace{2cm}
\includegraphics[width=0.80\textwidth,clip=true,angle=0]{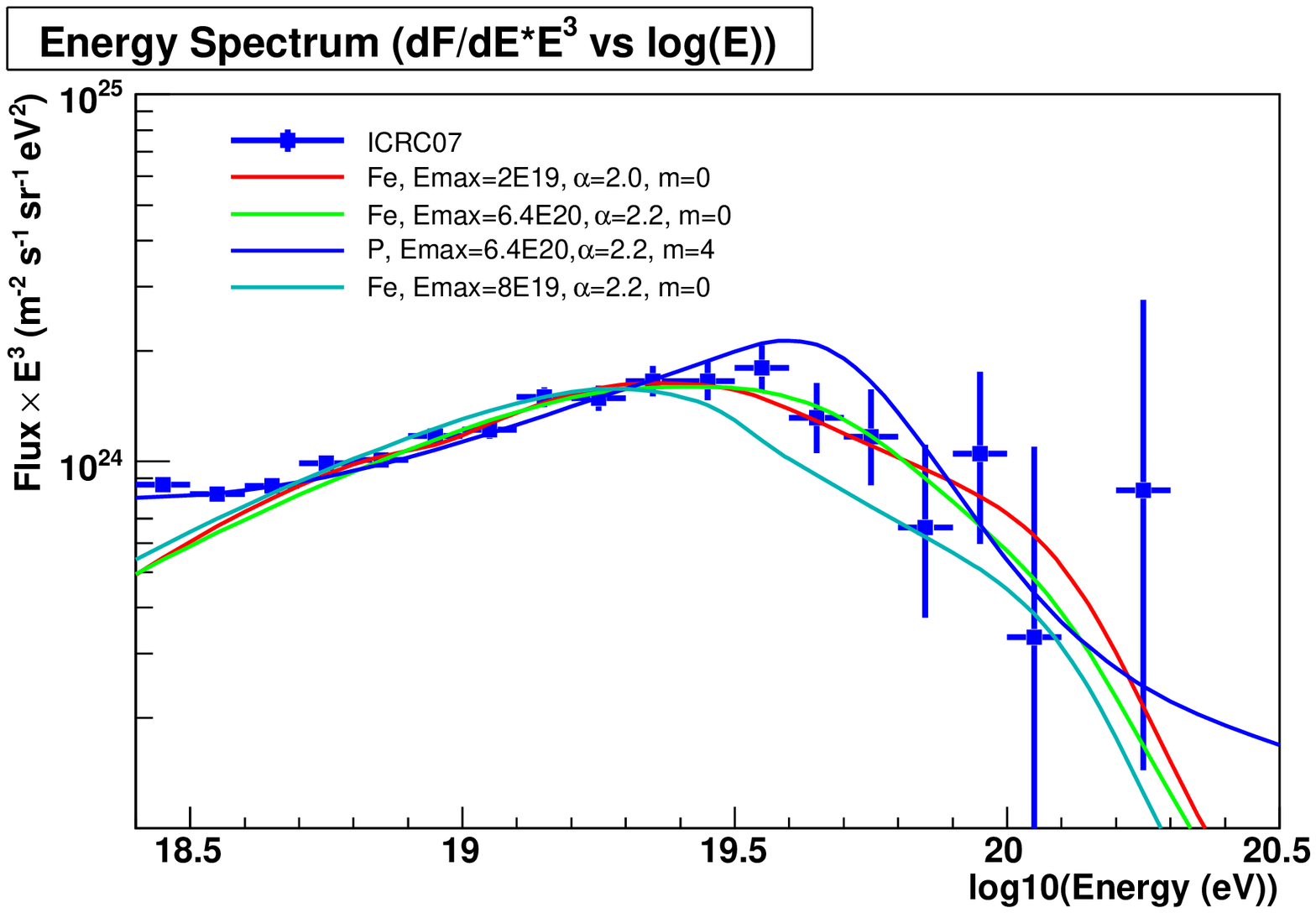}
\caption{Examples of predicted UHECR spectra compared to the Auger data for several models providing  good fits for p (blue),  low $E_{\rm max}$  Fe (red)  and high $E_{\rm max}$  Fe injected (green) and a bad fit (the
  intermediate $E_{\rm max}= 8 \times 10^{19}$ eV, $\alpha=2.0$, $m=0$ Fe case, in teal). The respective $p$-values of these models are: 0.119, 0.816, 0.744 and 0.0025. Recall that the maximum energy is $Z E_{\rm max}$.}
\label{FitExamples} 
\end{figure}

When $m=4$, good models with  pure proton injection have $\alpha=2.2$ and $E_{\rm max}=10^{20.2} -10^{21.1}$~eV  if $z_{\min} =0.000$ or $E_{\rm max}=10^{20.5} -10^{21.1}$~eV if $z_{\min} =0.005$. If all sources are further than 50 Mpc  ($z_{\min} =0.010$)  there are no good fits with proton injection, because the GZK cutoff becomes too sharp so the flux is too low at $E \leq 10^{20}$ eV. 
As $m$ decreases there are relatively more sources near by,  thus the initial energies are less redshifted and the sources contribute less to the spectrum at lower energies. This change is compensated in the models providing good fits by an increase in $\alpha$ (a steeper initial spectrum). However this  leads to too large a flux at energies below $E_{\rm cut}$ and the models are rejected by the low energy constraint. For $m=0$, for example, we see in  Fig.~ \ref{p-values-filter-Emax-alpha-mzero} that a good fit region allowed by the low energy constraint exists only for $\alpha=2.4$ and $E_{\rm max}=10^{21.1}$~eV and 
 $z_{\min} =0$. For large $E_{\rm max}$ the proton accumulation below the GZK energy increases, thus the normalization of the predicted flux need to be lower to provide a good fit,  this also lowers the flux predicted at low energies and the model is accepted by the low energy constraint.  
 For lower values of $\alpha$ the $p$-values of the pure proton injection  models are low because the predicted flux becomes too low at low energies, above but close to 
 $E_{\rm cut}= 1\times 10^{19}$ eV. This conclusion could be avoided if there was a non-negligible contribution from the LEC still contributing to the spectrum at  energies above $1\times 10^{19}$ eV.

 No model with pure oxygen injection provides a 
 good fit (outside the cross hatched region) with $m=4$ because for those models allowed by the low energy constraint, the protons resulting from the spallation of  the O nuclei produce a too large bump in the predicted spectrum at low energies above but close to $E_{\rm cut}$. However, for $m=0$ a high $E_{\rm max}$ region of good fits is  present for pure O injected,  with $\alpha=2.0$,  $E_{\rm max}>10^{20.5}$~eV if $z_{\min} = 0$ or 0.05, which disappears for larger distance to the sources.

Again for $m=4$, pure iron injection only provides acceptable models if the sources are close by, i.e.
$z_{\rm min}=0$ if  $\alpha=1.5-1.6$, and $E_{\rm max}=10^{20.2} -10^{20.5}$~eV.
  In contrast,  $m=0$ leads to a
larger region of satisfactory fits for iron injection which for $z_{\rm min}=0$ runs across all the range of
 $E_{\rm max}$ with $\alpha$ from 1.7 to 2.2. As the minimum distance to the sources increases to 25 Mpc  ($z_{\rm min}=0.05$) two separate regions of good fits remain for Fe injection: one at high  $E_{\rm max}$  and   $\alpha=1.9-2.2$ and one at low  $E_{\rm max}$ with $\alpha=1.7-2.0$. As the minimum distance to the sources increases to 50 Mpc  ($z_{\rm min}=0.1$) the low $E_{\rm max}$ region of good fits disappears and the high $E_{\rm max}$ region shrinks to a single combination of parameters, $E_{\rm max}=10^{21.1}$~eV,  $\alpha= 2.0$ (and it is clear that no good fit would remain if the sources would be even further away). Recall that the actual maximum energy of nuclei is $Z E_{\rm max}$.

\begin{figure}
\includegraphics[width=0.50\textwidth,clip=true,angle=0]{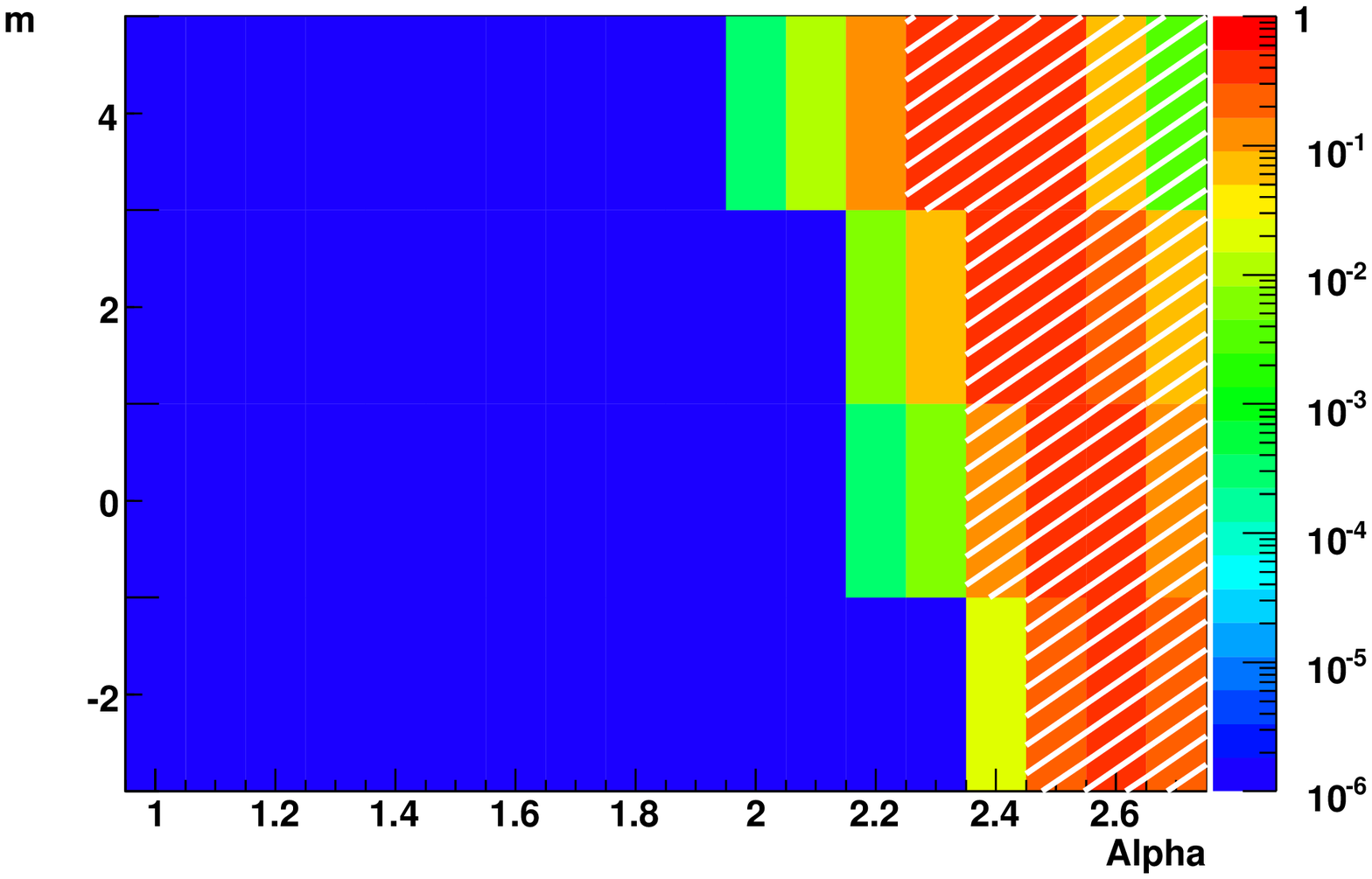}\hfill
\includegraphics[width=0.50\textwidth,clip=true,angle=0]{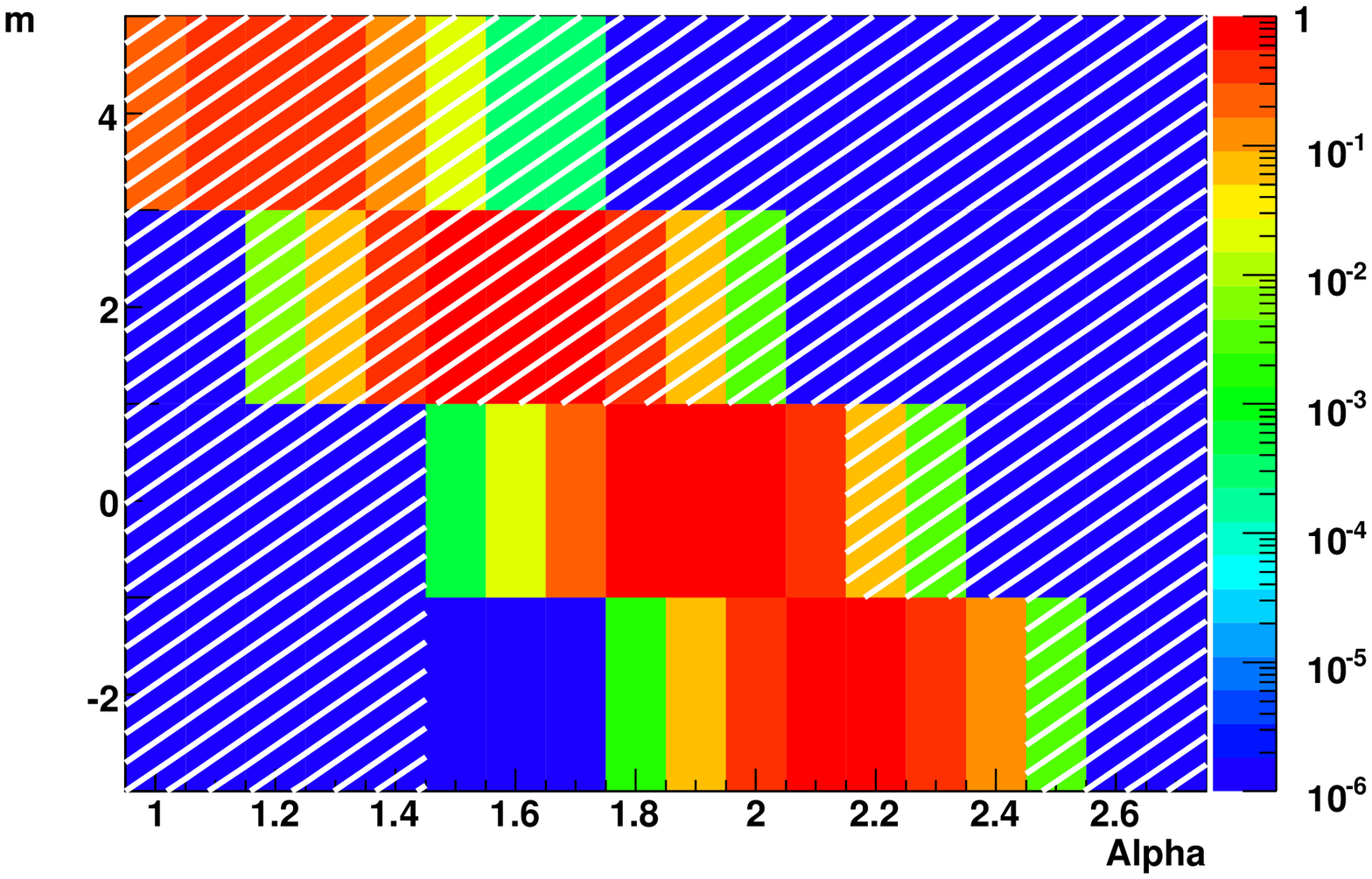} \hfill
\includegraphics[width=0.50\textwidth,clip=true,angle=0]{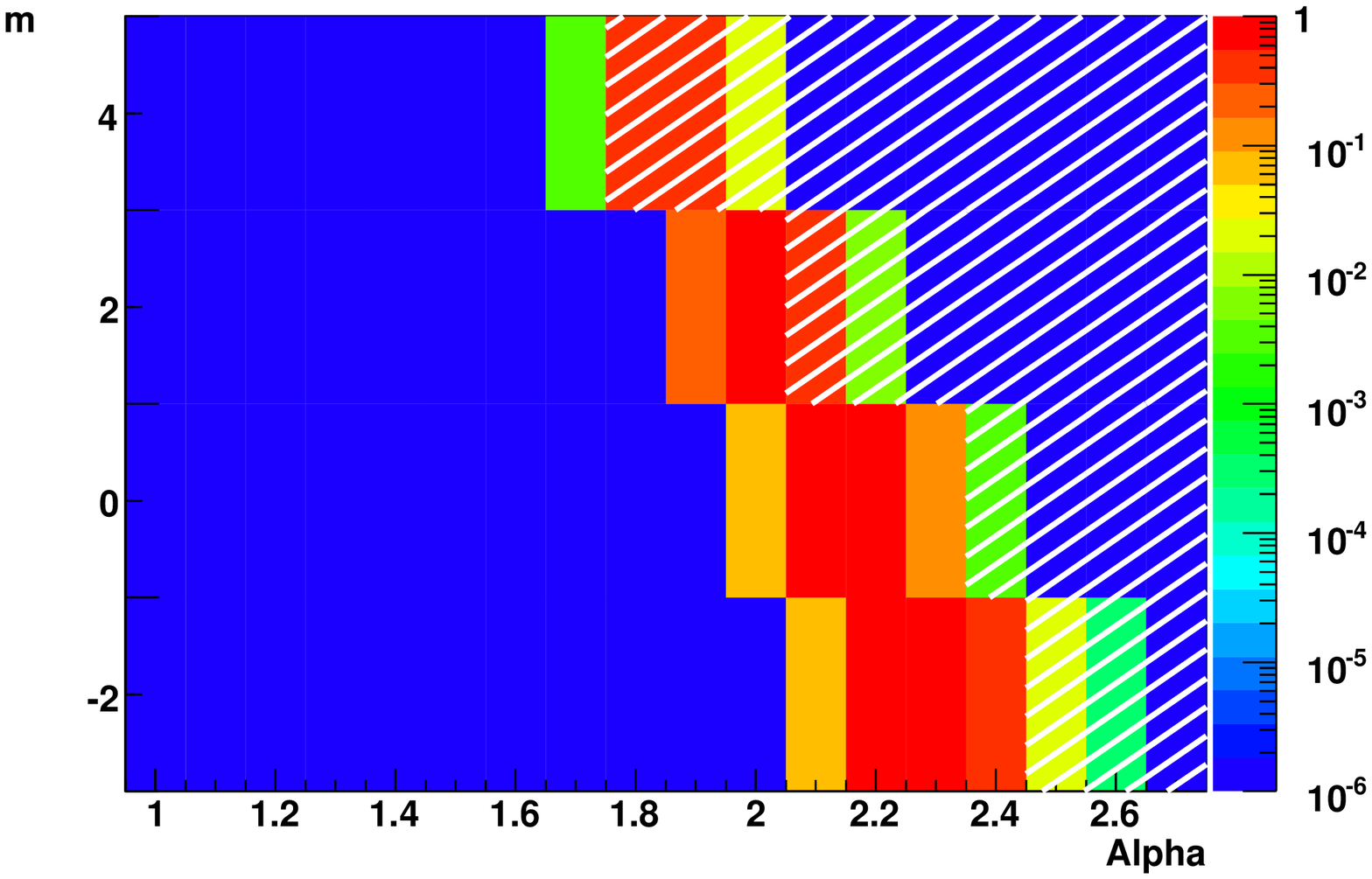} \hfill
\includegraphics[width=0.50\textwidth,clip=true,angle=0]{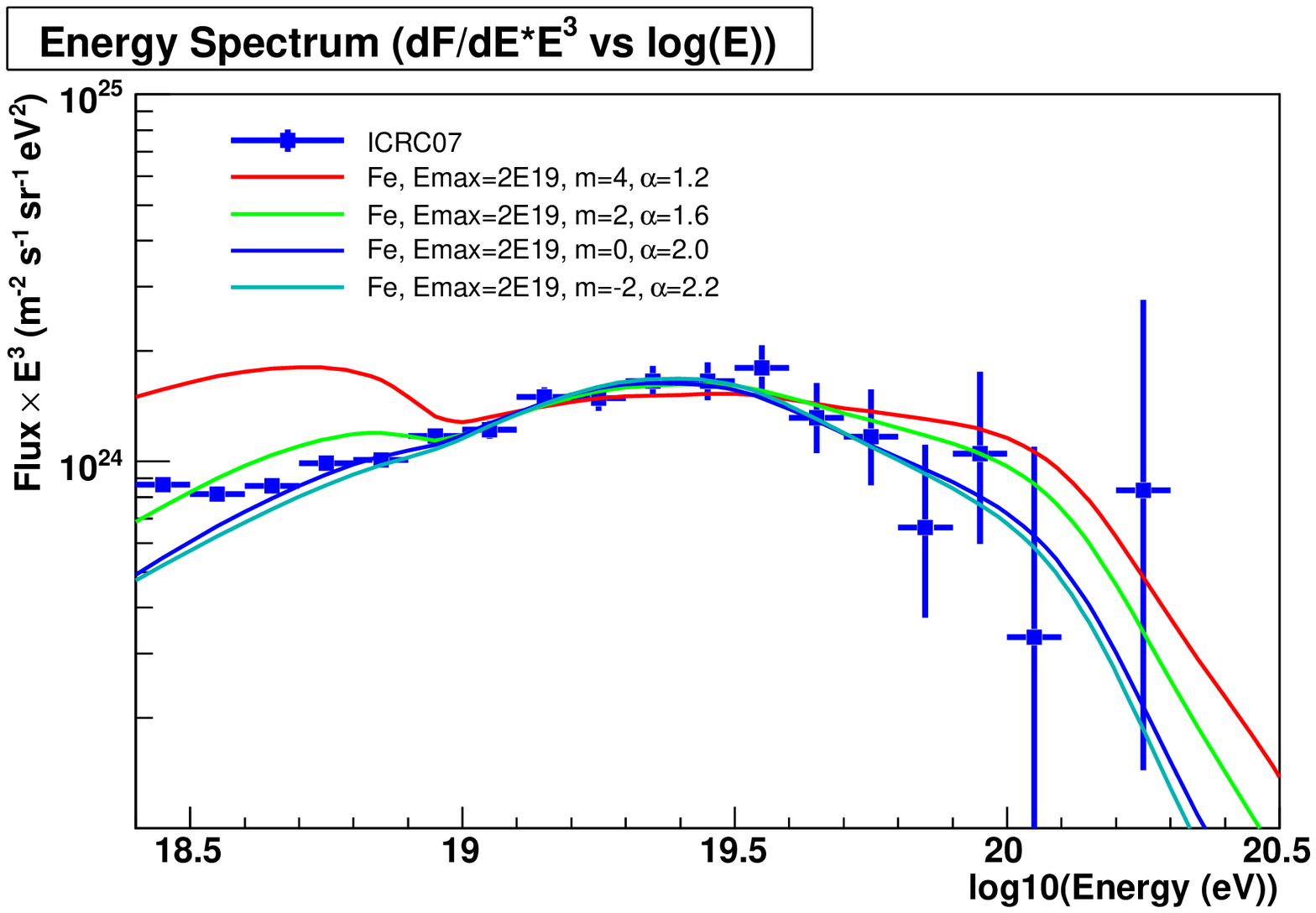} \hfill
\caption{Degeneracy in $m$ and $\alpha$ for p injection with $E_{\rm
    max} = 6.4 \times 10^{20}$~eV (top left), Fe injection with  maximum energy
     $Z E_{\rm max}=26 \times 2 \times 10^{19}$~eV (top right) and
  $26 \times 6.4 \times 10^{20}$~eV (bottom left). Predicted spectra for 
  best fit cases as function of $m$ for Fe injection with $E_{\rm max}=
 2 \times 10^{19}$~eV (bottom right). The $p$-values of the models listed are 0.458 ($m=4$), 0.855 ($m=2$), 0.816 ($m=0$) and 0.828 ($m=-2$).
}
\label{p-values-filter-m-alpha} 
\end{figure}

Closer sources, $z_{\min} = 0$, always provide better fits,
irrespective of the $m$ value, thus  in  the following  figures (Fig.~\ref{FitExamples} to
\ref{p-values-composition})  $z_{\min}$ is set to zero.  A few examples of the predicted  spectra  of models which provide good fits (i.e. having $p \ge 0.05$)  or bad fits are shown Fig.~\ref{FitExamples}. 

In Fig.~\ref {p-values-filter-m-alpha}  the degeneracy in $m$ and $\alpha$ is
shown for  three  of the models providing good fits,
    p injected with $E_{\rm max} = 6.4 \times 10^{20}$~eV (top left), Fe injected with  maximum energy
     $Z E_{\rm max}$  either $26 \times 2 \times 10^{19}$~eV (top right)  or
  $26 \times 6.4 \times 10^{20}$~eV (bottom left). The figure shows clearly that decreasing $m$ while increasing $\alpha$ yields the same results. The bottom right panel shows the predicted spectra for 
  best fit cases as function of $m$ from $-2$ to 4, for Fe injection with $E_{\rm max}=
 2 \times 10^{19}$~eV. 
 
 As we clearly see, requiring that the predicted flux does not exceed the observed flux below $E_{\rm cut}$ and above $2.5\times 10^{18}$ eV (the hatched regions in Figs.~\ref{p-values-filter-Emax-alpha} to \ref {p-values-filter-Ecut}) excludes a large number of otherwise good fits. Thus, the caveat we mentioned earlier against this constraint is relevant: the constraint would not hold if the extragalactic cosmic rays with  energy below some threshold energy between $2.5\times 10^{18}$ eV and  $E_{\rm cut}$ somehow do not reach Earth (are not emitted at the sources). 

 The best fits for proton injection happen for larger values of $\alpha$ (steeper spectrum) in comparison to the best fits for iron injection. As mentioned above, the steeper spectrum for p injection results in excess flux at  low energies, whereas the harder spectrum for Fe injection tends to give a deficit of flux at  low energies. As $m$ decreases (there are relatively more sources nearby), in order to get a good fit $\alpha$ must increase to compensate having less particles at low energy (close but above $E_{\rm cut}$) which means excess flux at  energies
  $E<E_{\rm cut}$. This means that only  large values of $m$ give acceptable solutions for proton. 
  
 The best fits for iron and oxygen, on the other hand, are forbidden by the low energy constraint due to an excess of flux at $E<E_{\rm cut}=1 \times 10^{19}$ eV due to  a bump consisting of protons produced by photodisintegration  (see the red Fe spectrum example for  $m=4$  in the bottom right panel of Fig.~\ref{p-values-filter-m-alpha}).  For larger $m$ values, good fits  require smaller $\alpha$ values which result in a larger flux at the higher energies which also means more photodisintegrated protons.

\begin{figure}[ht]
\includegraphics[width=0.50\textwidth,clip=true,angle=0]{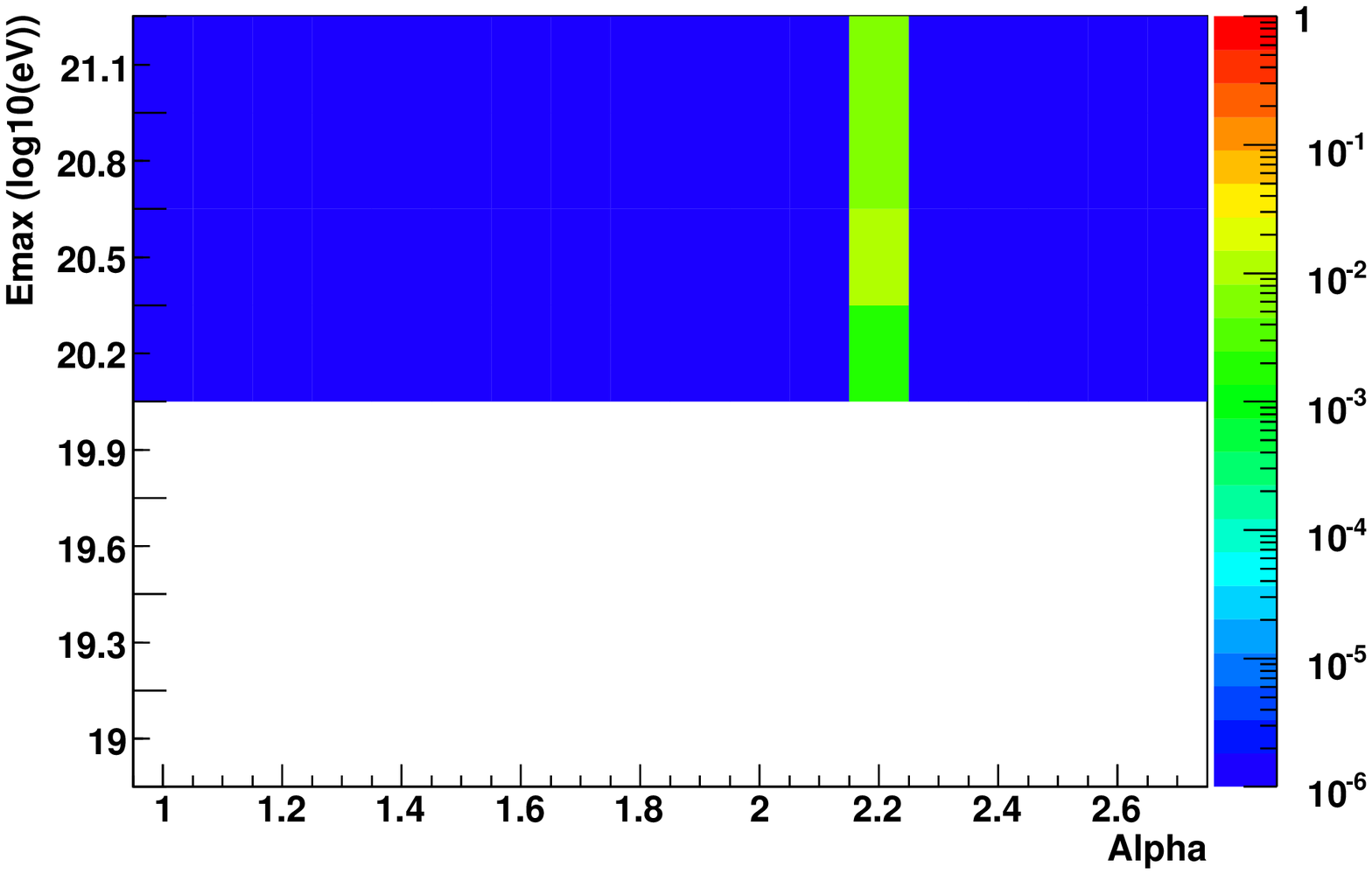}\hfill
\includegraphics[width=0.50\textwidth,clip=true,angle=0]{P_Zmin0.000EmaxAlpha_m4.eps} \hfill
\includegraphics[width=0.50\textwidth,clip=true,angle=0]{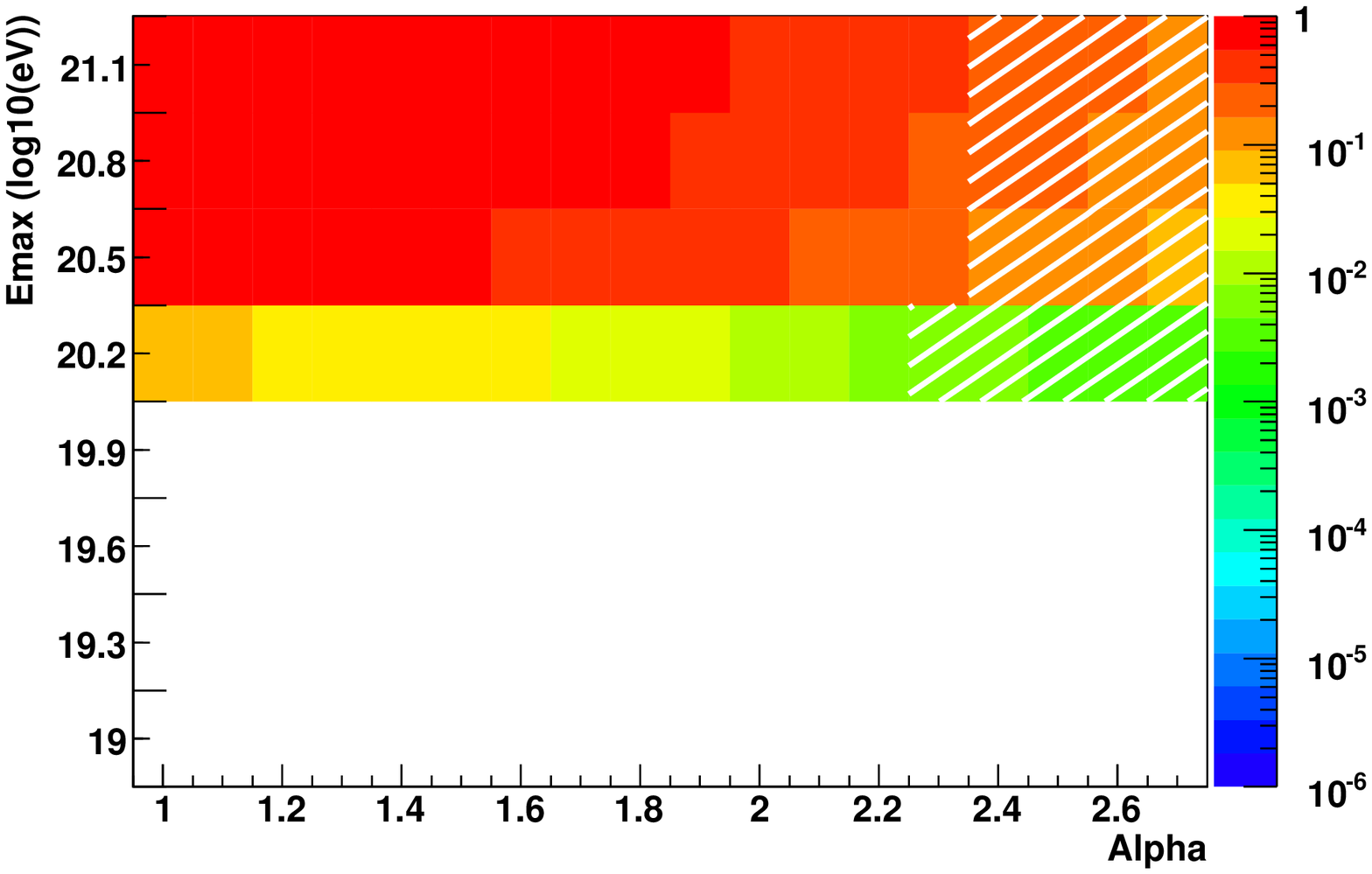} \hfill
\includegraphics[width=0.50\textwidth,clip=true,angle=0]{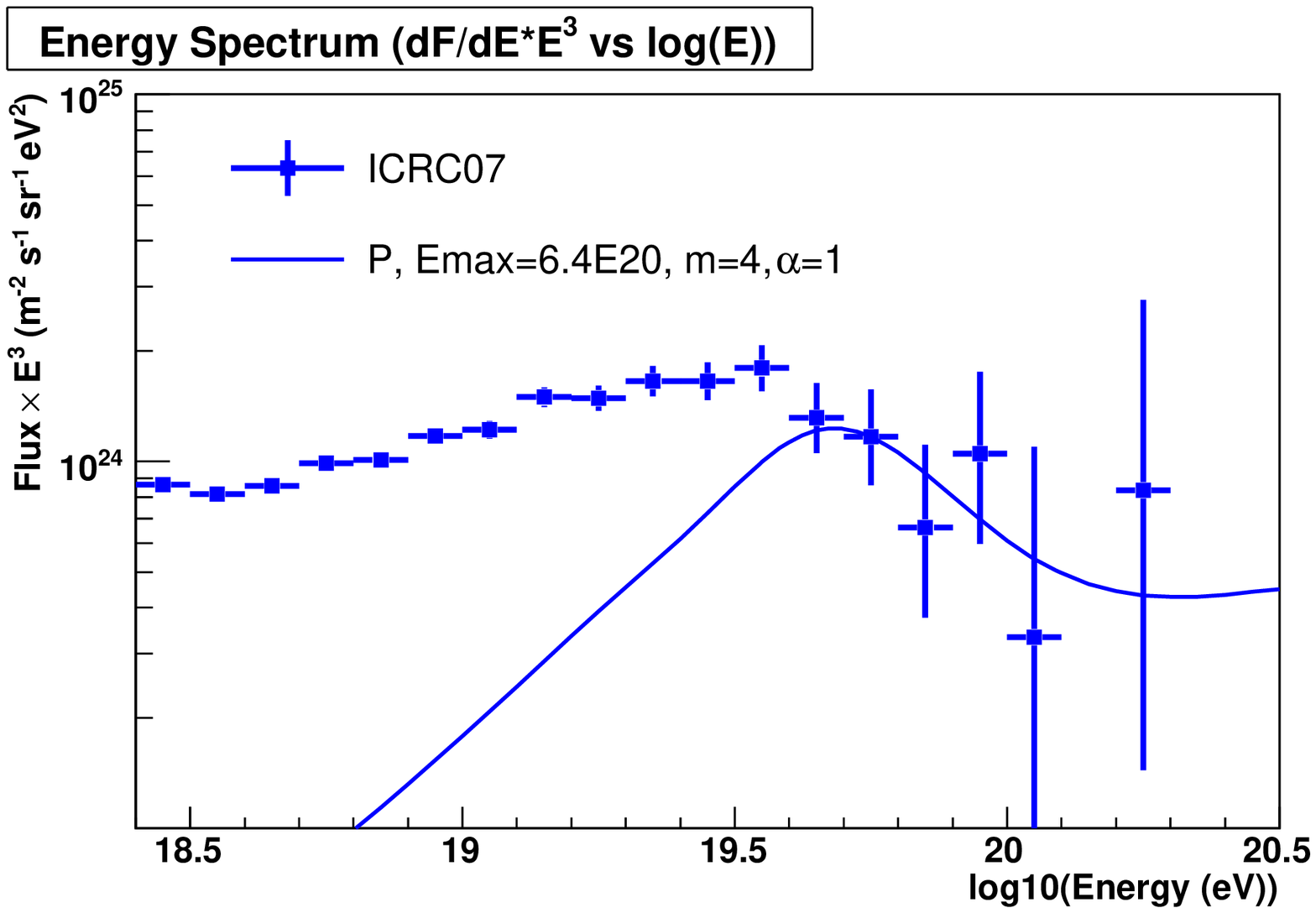}\hfill
\caption{Color coded $p$-value plots for  $z_{\min} = 0$ and $m=4$ and only protons injected
for different $E_{\rm cut}$ values:
 $2.5 \times 10^{18}$~eV (top left), $1 \times 10^{19}$~eV
  (top right), and $4 \times 10^{19}$~eV $E_{\rm cut}$ (bottom
  left). Example of the predicted flux  for $E_{\rm cut}=4 \times 10^{19}$~eV
  with $\alpha = 1$, $E_{\max} = 6.4 \times 10^{20}$~eV (bottom right), model with $p$-value 0.809.  }
\label{p-values-filter-Ecut}
\end{figure}

So far we have fitted the data above  $E_{\rm cut}=1 \times 10^{19}$~eV.
In Fig.~\ref{p-values-filter-Ecut} we explore the changes in the fits due to  different choices
of  $E_{\rm cut}$, namely  $2.5 \times 10^{18}$~eV and
$4 \times 10^{19}$~eV besides
 $1 \times 10^{19}$~eV for the case of proton
sources with $m=4$. 
 As mentioned earlier, each $E_{\rm cut}$ is
appropriate for different hypotheses for the energy at which the
transition to extra-galactic sources occurs. The effect of $E_{\rm cut}$ on the
goodness of fit is shown in Fig.~\ref{p-values-filter-Ecut}: 
the regions with acceptable $p$-values increase 
progressively with increasing $E_{\rm cut}$. This is easily understood, since
there are more events per bin at low energies, thus the error bars are smaller and  fewer models provide a good fit   for lower $E_{\rm cut}$.

 For $E_{\rm cut} = 2.5
\times 10^{18}$~eV,  the point $\alpha=2.2$, $E_{\rm max} = 10^{20.5}$~eV provides the best  fit although with $p < 0.05$.  If the first data bin, at the $10^{18.4}$~eV, is eliminated from the fit, the $p$-value becomes larger than 0.05. This is because the models with non-negligible $p$-value for this low $E_{\rm cut} $
 have a deficit of flux at the $10^{18.4}$~eV bin, the bin which has the smallest error bar. So presumably, if an LEC is added to match the flux exactly at that first bin, their low goodness of fit  could be improved.

 Fitting the spectrum only above $4 \times 10^{19}$~eV, on the other
hand, is easier and models with a wide range of $\alpha$ and
$E_{\rm max}$ values provide good  fits, especially for small  values of $\alpha$. 
Obviously these models require an LEC that
  makes up for the deficit in the flux below $4 \times 10^{19}$~eV. Even models with  $\alpha=1$,  a very flat spectrum, provide  good fits.
  With such a hard injection spectrum the flux below $4
\times 10^{19}$~eV is well under that of the Pierre Auger Observatory, as
shown in the bottom right plot of Fig.~\ref{p-values-filter-Ecut}.  We
can see from Fig.~\ref{p-values-filter-Ecut} that if the LEC is
assumed to extend all the way to $4 \times 10^{19}$~eV, almost any
combination of parameters is satisfactory.
 
Although Fig.~\ref{p-values-filter-Ecut} only shows the case of
proton injection with $m=4$, the same general consideration apply to nuclei and other values of $m$ as well (although, as we mentioned earlier, we cannot extend the fit all the way down to $2.5 \times 10^{18}$~eV for nuclei since we do not take magnetic deflections into account). 
In the following we use $E_{\rm cut}=1 \times 10^{19}$~eV.

\begin{figure}
\includegraphics[width=0.50\textwidth,clip=true,angle=0]{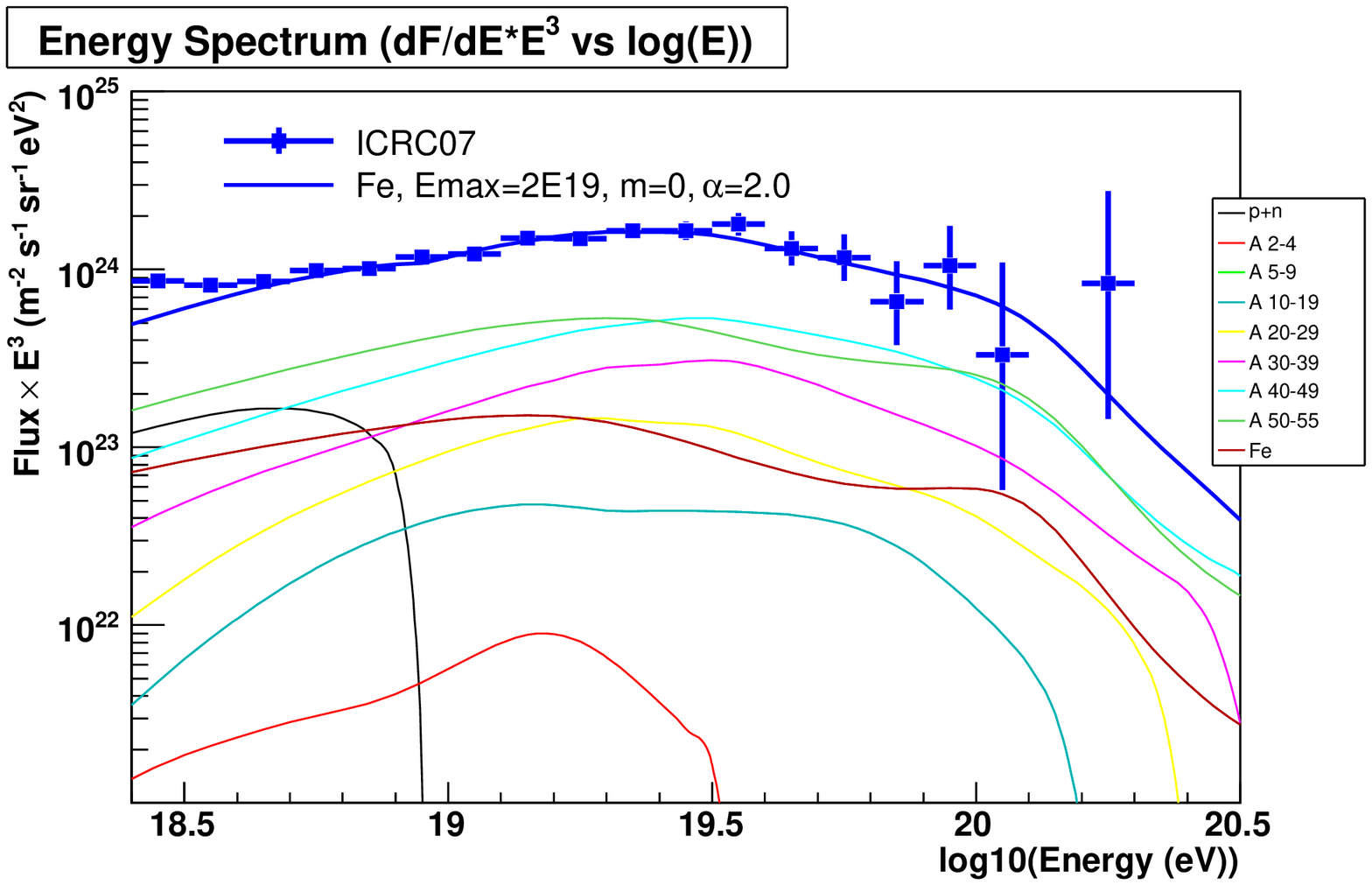}
\includegraphics[width=0.50\textwidth,clip=true,angle=0]{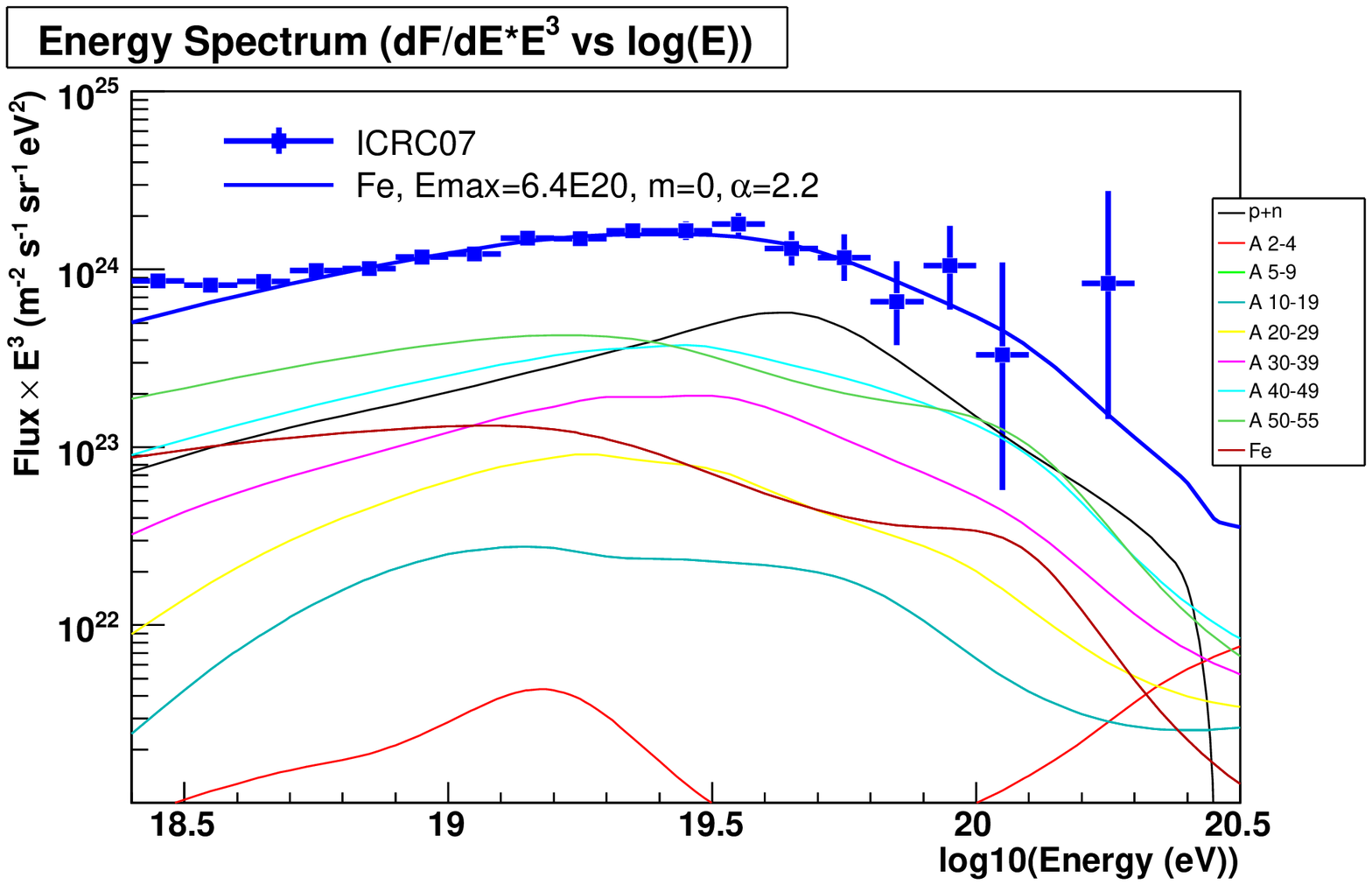}
\includegraphics[width=0.50\textwidth,clip=true,angle=0]{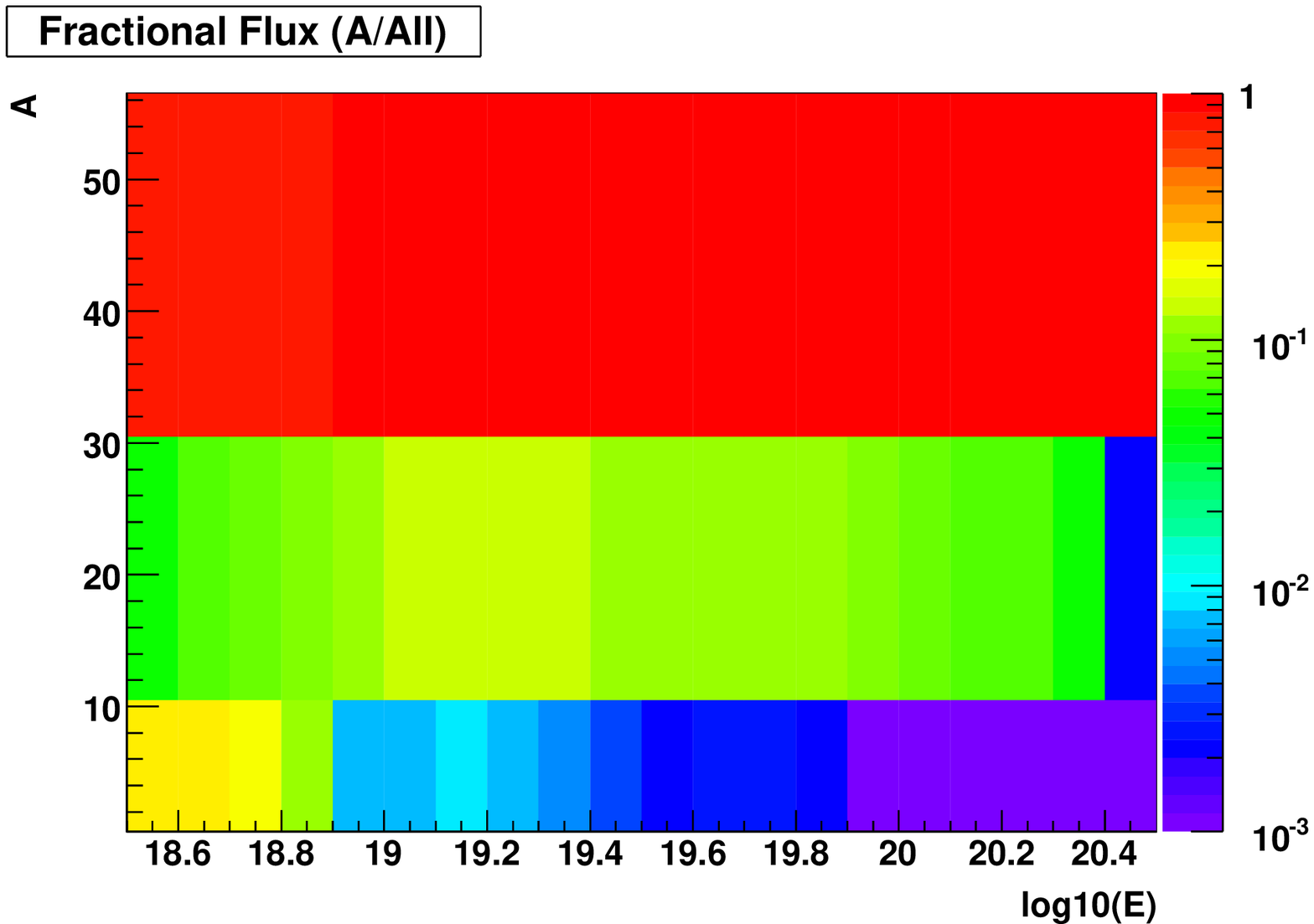}
\includegraphics[width=0.50\textwidth,clip=true,angle=0]{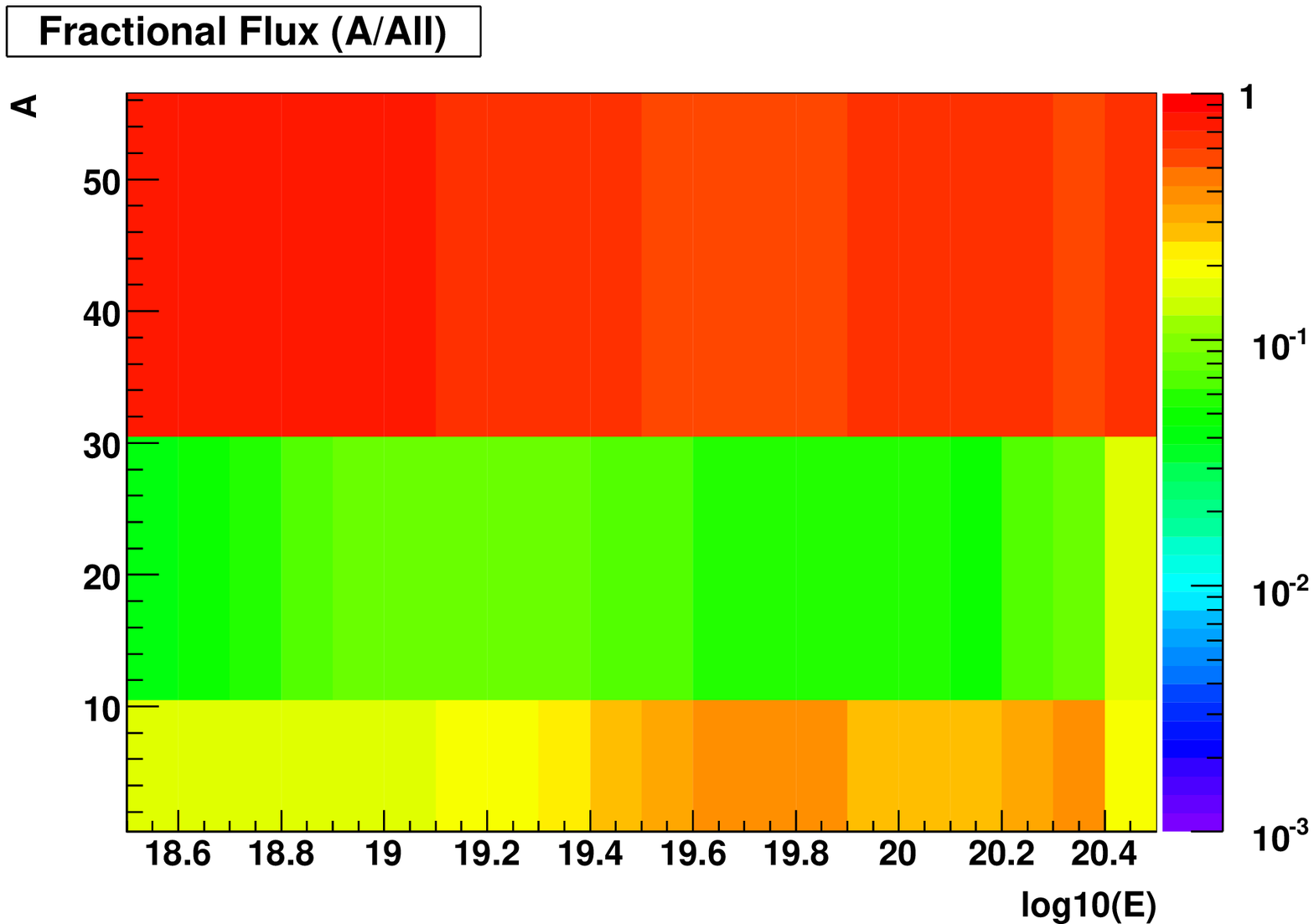}
\includegraphics[width=0.50\textwidth,clip=true,angle=0]{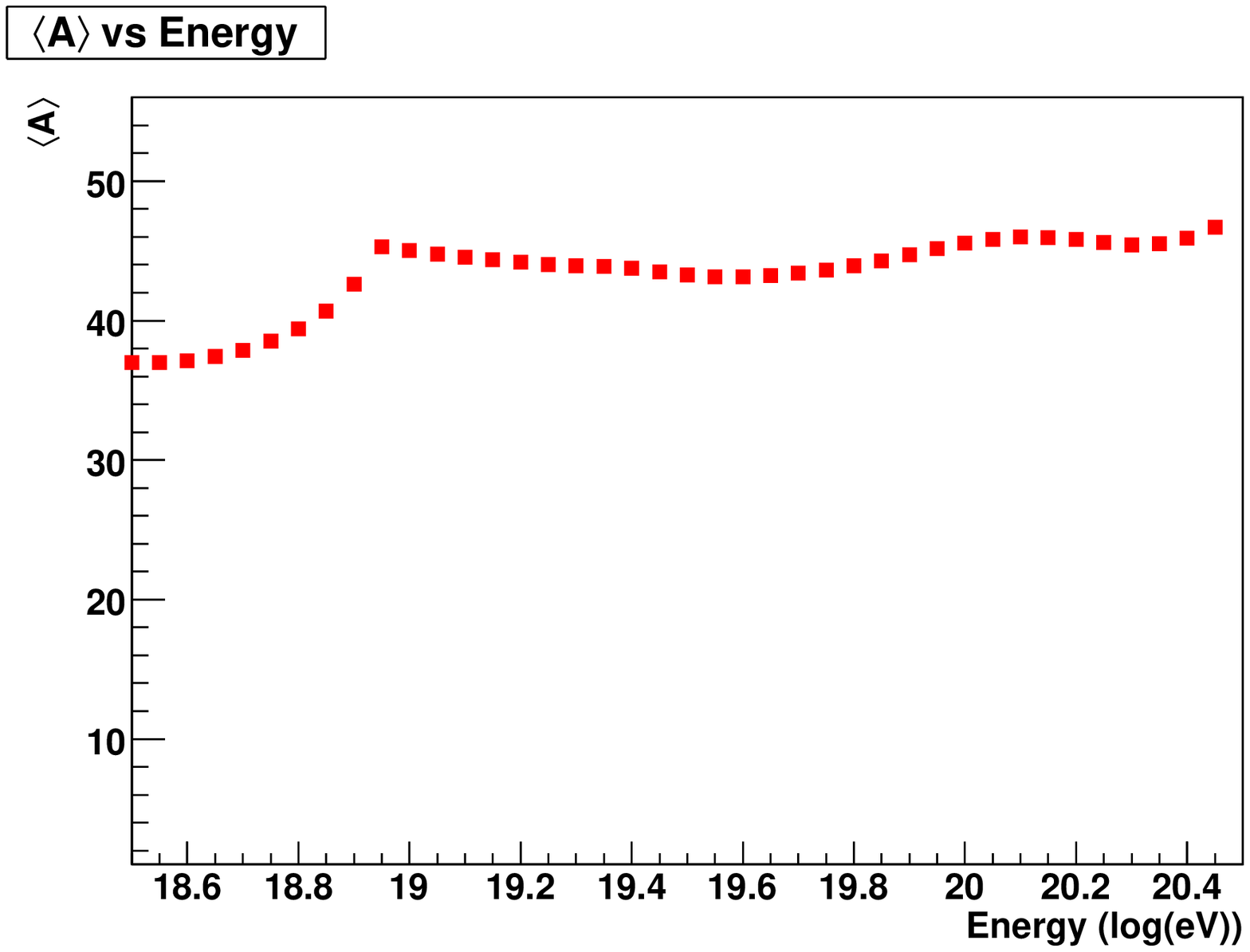}
\includegraphics[width=0.50\textwidth,clip=true,angle=0]{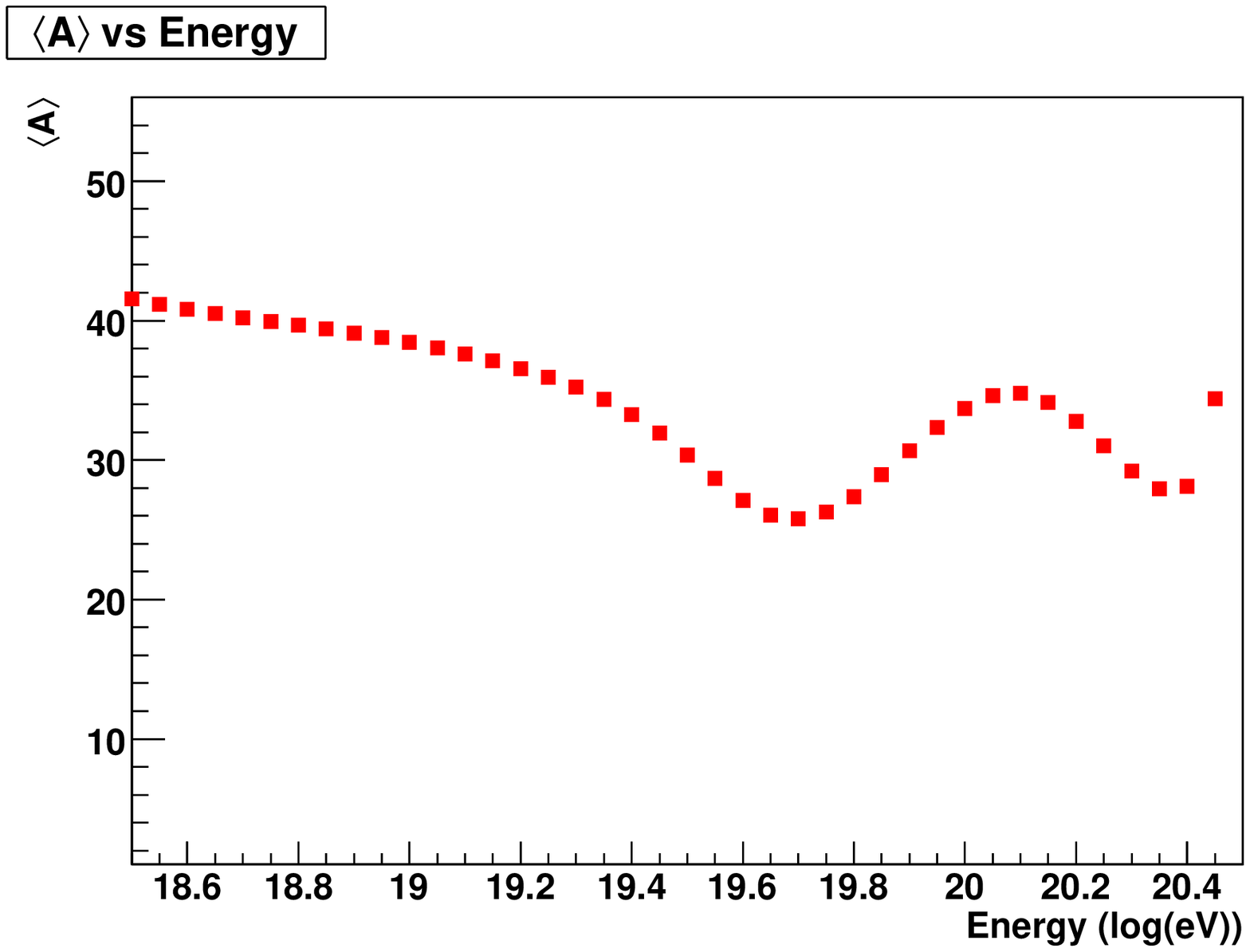}
\caption{Detailed breakdown of the UHECR composition from
  photodisintegration of pure Fe injected  for $Z E_{\rm max} = 26 \times 2 \times
  10^{19}$~eV (left column) and $Z E_{\rm max} = 26 \times 6.4 \times
  10^{20}$~eV  (right column) with $m=0$ ($\alpha=$ 2.0 and 2.2 respectively). Comparison of the flux of  different nuclei groups with the Pierre Auger spectrum (top panel) --- the
  black line is the flux of photodisintegrated protons. Fractional
  flux for different nucleus species, divided into 3 groups according to their atomic number A: from 1 to 10,  from 11 to 30, and  from 30 to 56, color coded according to the fractional flux
  (middle panels).  The flux is dominated by moderately heavy to heavy nuclei
  as indicated by the red regions, but  for $Z E_{\rm max} = 26 \times 6.4 \times 10^{20}$~eV it still contains a significant  proton flux  (middle right).  The light mass flux (A~=~1--10) is almost all
  protons and constitutes about 40\% of the total.  The average mass  as function of energy shown in
  the bottom panels.}
\label{p-values-composition} 
\end{figure}

\section{Composition and predicted spectra}

The two extreme best fit cases for iron with maximum energy  $ZE_{\rm max} = 26 \times 2
\times 10^{19}$~eV and $Z E_{\rm max} = 26 \times 6.4 \times
10^{20}$~eV have interesting implications for the composition of the
UHECR.  As explained above, if $m=4$ only the high $E_{\rm max}$ case provides good fits to the spectrum (and only if the sources are not farther away than 25 Mpc). Iron spallates during propagation and the breakdown of the
resulting composition is shown in Fig.~\ref{p-values-composition}. The
low $E_{\rm max}$ case is dominated by heavy elements with almost a
total absence of primaries lighter than Boron (atomic number five) throughout the entire
energy range. The proton flux from photodisintegration, the endpoint of
which is 1/56 of $Z E_{\rm max}$, ends below
$10^{19}$~eV. The high $E_{\rm max}$ case does contain a significant
proton fraction even up to the highest energies.  The average mass of
the composition for the low $E_{\rm max}$ case is upwards of 40~amu (see
bottom left in Fig.~\ref{p-values-composition}) nearing a flux of pure
iron above an energy of $1 \times 10^{19}$~eV.  The high $E_{\rm max}$
case has an average mass that varies with a minimum of 30~amu (see bottom right in
Fig.~\ref{p-values-composition}).   The detailed breakdown of the
compositional makeup of the UHECR flux is the key to distinguishing the two cases (see the two
 middle panels of  Fig.~\ref{p-values-composition} where the fraction of light, intermediate and heavy nuclei  is
represented in a color coded  logarithmic scale). Indeed, the
composition is dominated by the moderately heavy to heavy nuclei in
both cases, but  the fraction of   protons for the high E$_{\rm max}$
case is significant and can be upwards of 40\% (notice the orange line
at A~=~1~to~10 for all energies in the middle right plot).

 The latest data seem to  indicate  that the average X$_{\rm max}$, which  is
characteristic of a flux dominated by light elements below $2 \times
  10^{18}$~eV,  transitions to a value that reflects heavier
primaries especially above $\sim 2 \times
  10^{19}$~eV~\cite{ICRC2007-auger-composition}. If this turns out to
be true, it will be important to identify proton shower candidates in the
high energy regime.

The two types of Fe injection solutions providing a good fit to the Pierre Auger spectrum were also found, with a different statistical analysis and modeling of predicted spectrum, in a very recent paper
(see Ref.~\cite{Anchordoqui:2007fi}) in which also the   X$_{\rm max}$ data of Auger are used (they seem to fix $m=0$ and $z_{\rm min}=0$).

Very likely the UHECR sources will accelerate a mixed composition rather
than all iron or oxygen or protons.  We can think of a
scenario, then, where the endpoint of each nucleus species is $Z
\times 10^{19}$~eV at the source where $Z$ is the charge of the
nucleus. Then, the proton injection would end at a low energy, at 1/26 the maximum iron
energy, and the maximum energy 
for heavy nuclei would not be so high as to result in too
many protons from photodisintegration of heavy nuclei.  In this
scenario, unless the fraction of Fe (and other heavy elements) injected is very small,
we can have a mixed composition spectrum that is  dominated by heavy
elements at the highest energies. If, on the contrary, the endpoint
of each nucleus species is high, say $Z \times 10^{21}$~eV. Then presumably,
protons  would be the dominant component up to energies close to $10^{21}$~eV, since  hydrogen is the most abundant element.  The addition of the significant nucleon fraction from photodisintegrated heavy elements would only serve to strengthen the proton dominance.
A very recent paper (Ref.~\cite{Anchordoqui:2007fi}) addressed this issue using  injected mixtures of iron nuclei and protons for an $E_{\rm max}$ close to $4\times 10^{20}$~eV.  They found that for these energies a small component of a few percent iron still dominated the spectrum and composition at high energies.

\section{Conclusions}

 We have performed an exhaustive scan  in the source evolution factor $m$, the spectral index $\alpha$
and  maximum energy  $Z E_{\rm max}$ of the source spectrum and the minimum distance to the sources  $z_{\rm min}$,  for sources emitting only protons, or oxygen or iron  nuclei and compared the
total predicted flux at Earth above $E_{\rm cut} =1 \times 10^{19}$ eV with the latest Pierre Auger spectrum. We have also imposed the predicted spectrum not to exceed the observed one at energies below $1\times 10^{19}$ eV. For an evolution of sources with $m=4$, consistent with evolution of AGN, the  spectrum agrees with not only pure proton  injection (with $\alpha = 2.2$ and $E_{\rm max}=10^{20.2}- 10^{21.1}$  eV) but also  iron injection with (with $\alpha = 1.6-1.7$ and $E_{\rm max}= 10^{20.2}- 10^{20.5}$) if the sources are not further away than  50 and 25 Mpc respectively.
 
  For smaller $m$, in particular $m=0$, we find solutions with all injected compositions.
The iron injection is
particularly interesting in that it has two disparate regions of high
significance around $Z E_{\rm max} = 26 \times 2 \times 10^{19}$~eV (with $\alpha =1.7 -2.2$) and
$Z E_{\rm max} = 26 \times 6.4 \times 10^{20}$~eV (with $\alpha =2.0 -2.3$) with the intermediate
$E_{\rm max}$ cases much less favorable (only the high  energy solution remains if the sources 
are very far way). Our results for $m=0$ and  $z_{\rm min}$ seem to be in agreement with the results of
Ref.~\cite{Anchordoqui:2007fi}, which appeared while we were finishing writing the present paper.

We have also studied the effects
of $E_{\rm cut}$ and shown that the regions of parameter space with good fits depends strongly on it.
This is easily understood, since
there are more events per bin at low energies, thus the error bars are smaller and  fewer models provide a good fit   for lower $E_{\rm cut}$. Each $E_{\rm cut}$ is
appropriate for different hypotheses for the energy at which the
transition to extra-galactic sources occurs.  For $E_{\rm cut} = 2.5 
\times 10^{18}$~eV and pure proton injection, corresponding to the ``dip model" of Ref.~\cite{berezinsky2002}
only $\alpha=2.2$ provide models with a non negligible goodness of fit, with $E_{\rm max} = 10^{20.5}$~eV providing the best fit (although with $p < 0.05$). The almost good proton models for this low $E_{\rm cut}$ have a deficit of flux in the first fitted bin, at the $10^{18.4}$~eV bin, which has the smallest error bar. So presumably, if an LEC is added to match the flux exactly at that bin,
their goodness of fit would  improve. Also, if the first bin is eliminated from the fit, the best fit point just mentioned has $p\ge 0.05$.  This disagreement of the fit of the ``dip model" to the Auger spectrum using surface detector data due to the lowest energy bins, coincides with the recent findings of Berezinky~\cite{Berezinsky:2007wf} using a different statistical method.
 For  $E_{\rm cut}=4 \times 10^{19}$~eV,  good models are found regardless of $E_{\rm max}$, but a suitable low energy component should
become important up to energies close to $4 \times 10^{19}$~eV. 

The spectrum favors a minimum distance to sources, $z_{\min}$, that is as small as
possible and the degeneracy between $\alpha$ and $m$ was also demonstrated.

The three models that have the highest probability to describe the
observed spectrum paint very different pictures of cosmic ray
composition.  For the pure proton injection at the source  all UHECR primaries should be protons
(possible with some GZK photons), while the two iron injection cases lead to a mixed composition
separable by a distinctive abundance of UHECR proton primaries. The low $E_{\rm max}$ Fe injection case predicts that $\sim$~90\% of the primaries above an energy of $1\times
10^{19}$~eV are elements with an atomic weight greater than 30~amu,
whereas the high $E_{\rm max}$ case contains a fraction of protons only
slightly smaller  than the total flux of elements with an atomic weight
greater than 30~amu.  In both cases the  average atomic weight would be
considered heavy.

  If the hint of a transition from light element
dominance to heavy element dominance in  the composition of UHECR above $2 \times 10^{19}$~eV seen in the latest results of the Pierre Auger Observatory turns out to
be true, then the highest energy cosmic rays are likely to contain a
large fraction of heavy elements.  Both the low $E_{\rm max}$ case and
the high $E_{\rm max}$ case present an intriguing scenario for a mixed
composition.  Pure iron injection at the sources is
unlikely, so if cosmic rays are of mixed composition with maximum energy at the source
of each nucleus species equal to $Z \times 10^{19}$~eV, then the low
$E_{\rm max}$ case results in a composition that becomes heavier with
energy until only iron primaries remain.  In the high $E_{\rm max}$ case the composition also
becomes heavier with energy, but should maintain a significant flux of protons well
beyond the GZK energy, coming both from the proton injection itself and from the photodisintegration
of the heavy elements.  This certainly presents an intriguing
direction for composition and spectrum studies in the future. 

\verb''\ack

The work of G.G and O.K.
 was supported in part by NASA grants NAG5-13399 and ATP03-0000-0057. 
G.G was also supported in part by the US DOE grant DE-FG03-91ER40662
Task C.  K.A, J.L and M.H were supported in part by US DOE grant DE-FG03-91ER40662
Task F. The numerical part of this work was performed at the computer
cluster of the INR RAS Theory Division, the ``Neutrino" cluster of the 
UCLA Physics and Astronomy Department, as well as the Pierre Auger
cluster of the UCLA Physics and Astronomy Department.


\vspace{0.3cm}
\begin{thebibliography}{99}


\bibitem{gzk}
K.~Greisen,
Phys.\ Rev.\ Lett.\  {\bf 16}, 748 (1966).
G.~T.~Zatsepin and V.~A.~Kuzmin,
JETP Lett.\  {\bf 4}, 78 (1966)
[Pisma Zh.\ Eksp.\ Teor.\ Fiz.\  {\bf 4}, 114 (1966)].



\bibitem{agasa}
M.~Takeda {\it et al.},
Phys.\ Rev.\ Lett.\  {\bf 81}, 1163 (1998);
see N.~Hayashida {\it et al.},
astro-ph/0008102,
for an update;  see also
{\sf http~://www-akeno.icrr.u-tokyo.ac.jp/AGASA/}.
 
\bibitem{hires}
R.~U.~Abbasi {\it et al.}  [High Resolution Fly's Eye Collaboration],
Phys.\ Rev.\ Lett.\  {\bf 92}, 151101 (2004);
see also {\sf http~://hires.physics.utah.edu/}.

\bibitem{hires_mono_spec}
R.~Abbasi {\it et al.}  [HiRes Collaboration],
  astro-ph/0703099.

\bibitem{Auger} Pierre Auger Observatory, http://www.auger.org.

\bibitem{ICRC2007-auger-energy-spectrum}
  M.~Roth {\it et al.} [Pierre Auger Collaboration],
  ``Measurement of the UHECR energy spectrum using data from the Surface Detector of the Pierre Auger Observatory,''
  Proc. 30\textsuperscript{th} ICRC, M\'{e}rida (2007).

\bibitem{ICRC2005-auger-energy-spectrum}
  P.~Sommers {\it et al.} [Pierre Auger Collaboration],
  Proc. 29\textsuperscript{th} ICRC, Pune (2005) {\bf 7}, 387.

\bibitem{ICRC2005-auger-acceptance}
  D.~Allard {\it et al.} [Pierre Auger Collaboration],
  Proc. 29\textsuperscript{th} ICRC, Pune (2005) {\bf 7}, 71.

\bibitem{ICRC2007-auger-energy-determination}
  M.~Ave {\it et al.} [Pierre Auger Collaboration],
  ``Reconstruction accuracy of the surface detector of the Pierre Auger Observatory,''
  Proc. 30\textsuperscript{th} ICRC, M\'{e}rida (2007).

\bibitem{agasa_spec}
  M.~Takeda {\it et al.},
  Astropart.\ Phys.\  {\bf 19}, 447 (2003)
  [astro-ph/0209422]. 

\bibitem{50Mpc} F.W. Stecker, Phys. Lett. {\bf 21}, 1016 (1968); S.~Yoshida and M.~Teshima, 
Prog. Theor. Phys. {\bf 89}, 833 (1993); F.~A.~Aharonian and J.~W.~Cronin, {Phys. Rev.}
{\bf D50}, 1892 (1994); J.~W.~Elbert and P.~Sommers,
{Astrophys. J.} {\bf 441}, 151 (1995);

\bibitem{40Mpc} F.~Halzen, R.~A.~Vazquez, T.~Stanev, and V.~P.~Vankov,
  Astropart. Phys., {\bf 3}, 151 (1995).
  
\bibitem{dolag2004}
K.~Dolag, D.~Grasso, V.~Springel and I.~Tkachev,
JETP Lett.\  {\bf 79}, 583 (2004)
[Pisma Zh.\ Eksp.\ Teor.\ Fiz.\  {\bf 79}, 719 (2004)]; and
JCAP {\bf 0501}, 009 (2005).

\bibitem{Sigl:2004yk}
G.~Sigl, F.~Miniati and T.~A.~Ensslin,
Phys.\ Rev.\ D {\bf 68}, 043002 (2003);
%
astro-ph/0309695;
%
Phys.\ Rev.\ D {\bf 70}, 043007 (2004);
astro-ph/0409098.

\bibitem{ICRC2007-hires-elongation}
  G.~Hughes {\it et al.} [HiRes Collaboration],
  ``A Measurement of the Average Longitudinal Development Profile of Cosmic Ray Air Showers from 10$^{17}$~eV to 10$^{20}$~eV,''
  Proc. 30\textsuperscript{th} ICRC, M\'{e}rida (2007).

\bibitem{berezinsky2002}
  V.~Berezinsky, A.~Z.~Gazizov and S.~I.~Grigorieva,
  hep-ph/0204357.
 

\bibitem{ICRC2007-auger-composition}
  M.~Unger {\it et al.} [Pierre Auger Collaboration],
  ``Study of the Cosmic Ray Composition above 0.4 EeV using the Longitudinal Profiles of Showers observed at the Pierre Auger Observatory,''
  Proc. 30\textsuperscript{th} ICRC, M\'{e}rida (2007).


\bibitem{galactic_magn_field}
 T. T.~Stanev,
  Astrophys.\ J.\  {\bf 479}, 290 (1997);
  G.~A.~Medina-Tanco, E.~M.~de Gouveia Dal Pino and J.~E.~Horvath,
  astro-ph/9707041;
  M.~Prouza and R.~Smida,
  astro-ph/0307165.


\bibitem{kks1999}
O.E.~Kalashev, V.A.~Kuzmin and D.V.~Semikoz,
astro-ph/9911035.
  Mod.\ Phys.\ Lett.\ A {\bf 16}, 2505 (2001);
O.E.~Kalashev Ph.D. Thesis, INR RAS, 2003.

\bibitem{Gelmini:2007jy}
  G.~B.~Gelmini, O.~Kalashev and D.~V.~Semikoz,
  arXiv:0706.2181 [astro-ph].

\bibitem{Stecker:2005qs}
F.~W.~Stecker, M.~A.~Malkan and S.~T.~Scully,
  Astrophys.\ J.\  {\bf 648}, 774 (2006).

  \bibitem{AS2}
V. S. Berezinsky, et al, {\it ``Astrophysics of Cosmic Rays.''}
(North-Holland, Amsterdam, 1990);
T.K. Gaisser, {\it ``Cosmic Rays and Particle Physics.''}
(Cambridge University Press, Cambridge, England, 1990).

\bibitem{AS1.5}
R.J. Protheroe, In {\it ``Topics in cosmic-ray astrophysics''},
ed. M. A. DuVernois,
Nova Science Publishing: New York, 1999, (astro-ph/9812055);
M.A. Malkov, Ap.J. {\bf 511}, L53 (1999);
K. Mannheim, R.J. Protheroe, J. P. Rachen,
Phys. Rev. {\bf D63}, 023003 (2001).


\bibitem{peaks}
  E.~V.~Derishev, F.~A.~Aharonian, V.~V.~Kocharovsky and V.~V.~Kocharovsky,
  Phys.\ Rev.\ D {\bf 68}, 043003 (2003).

\bibitem{mono}
A.~Neronov and  D.~Semikoz, 
New Astronomy Reviews {\bf 47}, 693 (2003);
A.~Neronov, P.~Tinyakov and I.~Tkachev,
  J.\ Exp.\ Theor.\ Phys.\  {\bf 100}, 656 (2005)
  [Zh.\ Eksp.\ Teor.\ Fiz.\  {\bf 100}, 744 (2005)]
  [astro-ph/0402132].

\bibitem{Berezinsky:2002vt}
V.~Berezinsky, A.~Gazizov and S.~Grigorieva,
astro-ph/0210095;
  Phys.\ Lett.\  B {\bf 612}, 147 (2005).
 
  \bibitem{dp90}
J.~S.~Dunlop and J.~A.~Peacock, MNRAS {\bf 247}, 19 (1990). 

\bibitem{De Marco:2005kt}
  D.~De Marco, T.~Stanev and F.~W.~Stecker,
  Phys.\ Rev.\  D {\bf 73}, 043003 (2006)
 

\bibitem{Inoue:2007kn}
  S.~Inoue, G.~Sigl, F.~Miniati and E.~Armengaud,
  astro-ph/0701167.

\bibitem{AGASA_clusters}
M.~Takeda {\it et al.},
Astrophys.\ J.\  {\bf 522}, 225 (1999).

\bibitem{HiRes_clusters}
R.~U.~Abbasi {\it et al.}  [HIRES],
Astrophys.\ J.\  {\bf 610}, L73 (2004).
[astro-ph/0404137].

\bibitem{agasa_hires}
R.~U.~Abbasi {\it et al.} [The High Resolution Fly's Eye Collaboration
                  (HIRES)],
astro-ph/0412617.

\bibitem{agasa_hires_ok}
H.~Yoshiguchi, S.~Nagataki and K.~Sato,
  Astrophys.\ J.\  {\bf 614}, 43 (2004);
M.~Kachelriess and D.~Semikoz,
  Astropart.\ Phys.\  {\bf 23}, 486 (2005).

\bibitem{sources} 
E.~Waxman, K.~B.~Fisher and T.~Piran,
Astrophys.\ J.\  {\bf 483}, 1 (1997);
S.~L.~Dubovsky, P.~G.~Tinyakov and I.~I.~Tkachev,
Phys.\ Rev.\ Lett.\  {\bf 85}, 1154 (2000);
Z.~Fodor and S.~D.~Katz,
Phys.\ Rev.\ D {\bf 63}, 023002 (2001);
H.~Yoshiguchi, S.~Nagataki, S.~Tsubaki and K.~Sato,
Astrophys.\ J.\  {\bf 586}, 1211 (2003)
[Erratum-ibid.\  {\bf 601}, 592 (2004)];
%
H.~Yoshiguchi, S.~Nagataki and K.~Sato,
Astrophys.\ J.\  {\bf 592}, 311 (2003);
P.~Blasi and D.~De Marco,
Astropart.\ Phys.\  {\bf 20}, 559 (2004).

\bibitem{EGRET}
P.~Sreekumar {\it et al.}  [EGRET Collaboration],
Astrophys.\ J.\  {\bf 494}, 523 (1998).

 \bibitem{Gelmini:2007sf}
  G.~Gelmini, O.~Kalashev and D.~V.~Semikoz,
  astro-ph/0702464.

\bibitem{Albuquerque:2005nm}
  I.~F.~M.~Albuquerque and G.~F.~Smoot,
  Astropart.\ Phys.\  {\bf 25}, 375 (2006).

  \bibitem{auger-sd-photon-limit}
  J.~Abraham {\it et al.} [Pierre Auger Collaboration],
  ``Upper Limit on the Cosmic Ray Photon Flux Above $10^{19}$~eV Using the Surface Detector of the Pierre Auger Observatory,''
  Submitted to Astrop. Phys. (2007).
  
    
\bibitem{Gelmini:2005wu}
  G.~Gelmini, O.~Kalashev and D.~V.~Semikoz,
  arXiv:astro-ph/0506128, version 3 of 10/1/07.


\bibitem{Fodor-K-R}  Z.~Fodor, S.~D.~Katz and A.~Ringwald,
  Phys.\ Rev.\ Lett.\  {\bf 88}, 171101 (2002)
  and 
  JHEP {\bf 0206}, 046 (2002).

\bibitem{statistics} 
S. Baker and R.D. Cousins, Nucl. Instrum. Methods {\bf221}, 437 (1984); Particle Data Group's Statistics Review (2004).

\bibitem{Anchordoqui:2007fi}
  L.~A.~Anchordoqui, H.~Goldberg, D.~Hooper, S.~Sarkar and A.~M.~Taylor,
  arXiv:0709.0734 [astro-ph].
  
\bibitem{Berezinsky:2007wf}
  V.~Berezinsky,
  arXiv:0710.2750 [astro-ph].


\end{thebibliography}
\end{document}